# Freestanding Thin-Film Materials

*Li Liu* [a,b], *Peixin Qin* [a,b,*], *Guojian Zhao* [a,b], *Zhiyuan Duan* [a,b], *Jingyu Li* [a,b], *Sixu Jiang* [a,b], *Xiaoyang Tan* [a,b], *Xiaoning Wang* [c,*], *Ziang Meng* [a,b,*], *Zhiqi Liu* [a,b,*]

[a] School of Materials Science and Engineering, Beihang University, Beijing 100191, China
[b] State Key Laboratory of Tropic Ocean Engineering Materials and Materials Evaluation, Beihang University, Beijing, 100191, China
[c] The Analysis & Testing Center, Beihang University, Beijing, 100191, China
* Corresponding authors.
*E-mail addresses:* E-mail: qinpeixin@buaa.edu.cn; wangxiaoning1997@buaa.edu.cn; mengza@buaa.edu.cn; zhiqi@buaa.edu.cn

**Keywords:**

freestanding thin films, physical delamination, sacrificial layer etching, transfer techniques, strain engineering

**Abstract:**

Freestanding thin films, a class of low-dimensional materials capable of maintaining structural integrity without substrates, have emerged as a forefront research focus. Their unique advantages—circumventing substrate clamping, liberating intrinsic material properties, and enabling cross-platform heterogeneous integration—underpin this prominence. This review systematically summarizes core fabrication techniques, including physical delamination (e.g., laser lift-off, mechanical exfoliation) and chemical etching, alongside associated transfer strategies. It further explores the induced strain modulation mechanisms, extreme mechanical properties and interface decoupling effects enabled by these films. Representative case studies demonstrate breakthrough applications in flexible/ultrathin electronics, ultrahigh-sensitivity sensors and the exploration of novel quantum states. Critical challenges regarding scalable fabrication, precise interface control, and long-term stability are analyzed, concluding with prospects for emerging applications in bio-inspired intelligent devices, quantum precision sensing, and brain-inspired neural networks.

# 1. Introduction

Freestanding thin films refer to ultrathin materials (ranging from nanoscale to microscale thickness) that maintain structural and functional integrity without the need for a supporting substrate [1,2]. The initial realization and formal introduction of freestanding films can be traced back to 1966 and 1978, with chemical exfoliation emerging as the earliest method for fabricating flexible films [3,4]. In early studies, the concept of "freestanding" primarily denoted the geometric independence of the material from a substrate, characterized by a broad thickness range (micrometers to millimeters), where mechanical stability was maintained by macroscopic structural design. This definition was widely adopted in subsequent research following the 1990 GaAs nanowire studies [5-8]: examples include InP membranes formed via bonding-and-release processes [9], and thicker films such as nanosheets/fibers or polymer membranes produced by spin-coating [10,11] or electrospinning [12-14]. These fabrication strategies were primarily aimed at achieving transferability, but they overlooked two critical aspects: the evolution of interfacial stresses during heterogeneous integration, and the inherent bulk-like properties exhibited by the materials due to their thickness limitations, rendering them fundamentally distinct from truly ultrathin systems.

With the rapid advancement of nanotechnology and advanced devices, research on freestanding thin films has garnered unprecedented attention, emerging as a pivotal frontier for overcoming performance limitations in conventional materials and devices. Traditional thin films, constrained by substrate lattice matching, thermal expansion coefficient mismatch, and interfacial effects (termed the "substrate clamping effect"), often exhibit suppressed or distorted intrinsic physicochemical properties (e.g., ferroelectricity, ferromagnetism, superconductivity, mechanical flexibility, optical response) [15-24].

The core value of freestanding films lies in their "substrate-free" nature—fully liberating the intrinsic properties of materials at low-dimensional limits while providing revolutionary design freedom for material engineering and device architecture. These films can be precisely transferred, stacked, folded, or curled as independent units, enabling van der Waals (vdW) heterostructures [25,26], complex three-dimensional suspended architectures [27-29], and novel micro/nanoelectronic systems [30] unattainable via conventional epitaxy. They also achieve near-theoretical-limit mechanical strength, flexibility, sensing sensitivity, and mass transport/optical efficiency at nano-to-atomic-scale thicknesses [31-34]. These attributes establish freestanding films as fundamental building blocks for high-performance flexible electronics, ultra-sensitive sensors [35,36], ultrahigh-flux separation membranes [37], and advanced photonic devices [2]. Furthermore, the absence of substrate constraints opens new avenues for exploring exotic quantum states and phase transitions in strongly correlated quantum materials (e.g., high-temperature superconductors [38-40], and Mott insulators [41-44]).

Nevertheless, the fabrication and transfer of freestanding films remain critical challenges for constructing functional heterostructures and flexible electronics. Their separation techniques follow an evolutionary trajectory rooted in two-dimensional (2D) material research. Despite significant differences in material dimensionality and bonding strength, core methodologies share substantial similarities. Mechanical exfoliation, first demonstrated for isolating monolayer graphene [45] and later extended to layered materials (e.g., h-BN [46], $WS_2$ [47,48], $NbSe_2$ [49]), now serves as a primary technique for freestanding film fabrication. The key distinction arises from intrinsic material properties: 2D material exfoliation relies on interlayer vdW cleavage, whereas freestanding film separation predominantly employs stressor-layer-induced interfacial strain release [50,51] or graphene-mediated "remote epitaxy" for vdW lift-off [52,53]. Strong interfacial coupling in conventional epitaxial films fundamentally limits mechanical exfoliation in preserving film integrity and crystallinity [54-56], necessitating precise control of residual stress relaxation and interfacial cleanliness. Sacrificial substrate etching represents a universal delamination approach—chemical vapor deposition (CVD) of 2D materials often employs metal foils (e.g., Cu) as sacrificial templates [57,58], aligning technically with sacrificial-layer etching for freestanding thin films [59].

Transfer methodologies exhibit even greater commonality: Polymer-assisted layers (e.g., rigid polydimethylsiloxane PMMA carriers or flexible Polymethyl Methacrylate PDMS stamps) are widely adopted for damage-free transfer in both systems. Standardized processes include "stamp transfer" (using elastomers like PDMS) and "wet transfer" (relying on hard supports like PMMA) [60-66]. A critical divergence exists: Freestanding films require external stabilization to mitigate thickness-induced brittleness and residual-stress-driven deformation [67,68], whereas 2D materials leverage atomic-thickness-enabled intrinsic flexibility and weak interlayer interactions for direct dry transfer onto target substrates [63,66].

Freestanding films, with their preserved structural integrity and superior properties, provide an advanced platform for designing and integrating functional materials. Recent years have witnessed intense focus on their fabrication and transfer—particularly for epitaxial single-crystalline films—spurring novel methods like laser lift-off [69,70], sacrificial-layer etching [15,54,58], and remote epitaxy van der Waals lift-off [52,71,72]. Among these, water-soluble sacrificial layers (e.g., SrO [73], La$_{0.7}$Sr$_{0.3}$MnO$_3$ [54], Sr$_3$Al$_2$O$_6$ [74], Sr$_4$Al$_2$O$_7$ [75]) have proven effective across diverse complex oxide systems. Their exceptional properties and emergent phenomena highlight transformative potential in strain engineering [76-78], multifunctional integration [15,26,79,80], emergent quantum effects [25,33], and extreme-environment adaptability [76,80].

This review comprehensively examines the evolution of freestanding thin films, encompassing fabrication techniques, transfer/integration strategies, performance optimization, and cross-disciplinary applications (Fig. 1). We further prospect their development trajectory and potential in frontier interdisciplinary fields.

## 2. Freestanding Thin Films Fabrication

Heteroepitaxy of high-quality thin films traditionally relies on rigid single-crystal substrates (e.g., sapphire, SiC) for lattice matching, yet faces three fundamental constraints: lattice mismatch-induced high-density dislocations [81]; limited thermal management efficiency due to substrate thermal conductivity [82,83]; and high cost [84]. Concurrently, flexible integration presents intrinsic challenges: direct growth of oxide films on polymer (PI/PET, etc. [85-87]) or metal foil (Ni/Cr, etc. [88-90]) substrates typically yields amorphous/polycrystalline structures, failing to meet the atomic-level interface ordering required for heteroepitaxy. To reconcile intrinsic properties (e.g., high carrier mobility, ferroelectric ordering) with flexibility demands, film fabrication and transfer technology has emerged as a critical solution. This involves: (i) growing high-quality epitaxial films on single-crystal substrates, (ii) removing substrates via physical delamination or chemical etching, and (iii) obtaining freestanding single-crystalline films for cross-platform functional integration (see Table 1) [28,50,74].

### 2.1 Physical Delamination Techniques

Epitaxial films typically bond to substrates via strong chemical bonds, making mechanical delamination challenging except for rare spontaneously detachable systems [55]. Researchers have developed diverse physical delamination methodologies to obtain freestanding thin films. This section details representative techniques including laser lift-off, mechanical exfoliation, and remote epitaxy vdW lift-off.

**2.1.1 Laser Lift-Off**

Laser Lift-Off (LLO) is a pivotal technique for damage-free separation of freestanding thin films. Its fundamental principle relies on selective energy deposition by short-pulsed lasers at an absorption layer within the film-transparent substrate interface: The laser penetrates the transparent substrate and is efficiently absorbed by the interfacial layer, triggering instantaneous thermal energy conversion that induces violent material vaporization and decomposition. Rapidly expanding gaseous products generate high-pressure impact forces (microplasma/stress waves), leading to complete film detachment when exceeding the interfacial binding strength (Fig. 2a, b). This technique is particularly suitable for fabricating high-quality single-crystalline/polycrystalline freestanding thin films [91-93].

Since its landmark demonstration for GaN film detachment from sapphire substrates in the late 20th century, LLO has continuously advanced optoelectronic devices [83,84,94-96]. Its core mechanism involves interface-selective photothermal decomposition: The wide bandgap of sapphire permits ultraviolet laser transparency, while the narrow bandgap of interfacial GaN enables strong photon absorption. This induces transient high temperatures that decompose GaN into liquid gallium and nitrogen gas. Nitrogen expansion generates separation pressure, complemented by gallium's interfacial lubrication effect, enabling damage-free detachment via post-processing.

Early studies validated micromachining feasibility using 355-nm laser-induced thermal decomposition [94]. Subsequent advances with 248-nm excimer lasers and epoxy bonding achieved full transfer of millimeter-scale GaN films to silicon substrates, with X-ray diffraction confirming preserved crystallinity [84]. In 2016, a defocused 266-nm solid-state laser irradiation strategy enabled selective precision lift-off for micron-scale light-emitting diodes (LEDs) (Fig. 2a) [70]. In 2022, Sun *et al.* leveraged the energy match between 355-nm picosecond ultraviolet laser photons (3.5 eV) and the GaN bandgap (3.4 eV), inducing intrinsic absorption-plasma synergy at ultralow energies (0.09–0.13 J/cm$^2$) to strictly confine delamination at the GaN/sapphire interface [97].

Building on GaN systems, LLO has been extended to ferroelectric oxides (e.g., (Pb,La)(Zr,Ti)$O_3$ (PLZT), Pb(Zr,Ti)$O_3$ (PZT)). Selective laser irradiation enables transfer from rigid substrates (MgO, sapphire) to flexible carriers (PET, PI) while retaining ferroelectric/piezoelectric properties [35,36,95,98-100]. To meet ultrathin high-density flexible device demands, fabrication precision is enhanced by synergistic innovations including: (i) The introduction of organic sacrificial layers (e.g., polyimide PI [101], polyurethane PU [102,103]) or inorganic sacrificial layers (e.g., amorphous gallium oxide ($α$-GaO$_x$) [104], hydrogenated amorphous silicon (a-Si:H) [29,105]) significantly minimizes interfacial damage, enabling large-area lift-off of $Bi_2Te_3$ thermoelectric films and PLZT piezoelectric films as well as the fabrication of flexible Organic LED/resistance random access memory (RRAM) memory arrays (Fig. 2c, d). (ii) Replacing thermal decomposition with ion-implanted damage layers [69] or integrating $SiO_2$ buffers with continuous-wave lasers to suppress thermomechanical damage [29] (Fig. 2a) achieves breakthroughs in Si integration and high-temperature CMOS compatibility, laying foundations for complex architectures.

Laser lift-off offers several notable advantages. The selective photothermal effect within sacrificial layers generates delamination stress with minimal damage to the film, as exemplified by GaN systems, while maintaining high process selectivity. Moreover, transparent substrates such as sapphire can be recycled to lower fabrication costs, and the non-contact nature of the technique enables both large-area and patterned processing. Despite these merits, LLO still faces critical limitations. The process is highly sensitive, requiring precise control of laser parameters in addition to expensive equipment. Furthermore, residual defects often remain on the detached surfaces, necessitating post-treatments such as annealing. Its applicability is also restricted to configurations combining a transparent substrate with a strong absorber layer, where the high-pressure stress generated during processing may lead to cracking in ultrathin films.

**2.1.2 Mechanical Exfoliation/Spalling**

Mechanical exfoliation is a widely employed method for producing freestanding thin films, and its principle relies on two mechanisms. The first is the direct utilization of intrinsically weak interlayer bonding, such as van der Waals gaps in layered materials, to achieve exfoliation through applied physical forces, whereas the second takes advantage of the inherently weak adhesion between the film and the substrate, which serves as a predefined separation plane.

Taking layered FeSe as an example [39,40,106,107], researchers adapted graphene preparation techniques to mechanically exfoliate 10–100 nm flakes from bulk FeSe crystals using low-residue tapes or PDMS. However, air exposure induced selenium volatilization and iron oxidation, suppressing the superconducting transition temperature $T_S$ to 3–7 K (9 K in bulk). h-BN encapsulation and a nitrogen environment increased the residual resistance ratio (RRR) to >16 while preserving 2D transport properties. Similarly, in strongly anisotropic $Bi_2Sr_2CaCu_2O_{8+\delta}$ crystals, superconducting $CuO_2$ layers are isolated by insulating BiO/SrO layers. Mechanical cleavage via vdW forces between BiO layers yields single-crystalline flakes with thicknesses ranging from ~1.5 nm to hundreds of nanometers (minimum ~1.5–1.8 nm). These samples exhibit atomically flat surfaces and stable adhesion to $SiO_2$/Si substrates [108].

Mica is a common layered substrate for freestanding film fabrication [109-111]. For instance, $BaTiO_3$ (30 nm)/$SrRuO_3$ (100 nm) heterostructures grown on mica via pulsed laser deposition (PLD) [112] can be exfoliated into flexible films containing 20 μm mica, demonstrating reversible mechanical strain modulation of the Curie temperature $T_C$ and magnetic anisotropy (Fig. 3a). Although the exfoliated structure retains mica, the functional layer detaches from the bulk substrate, enabling transfer, bending, and attachment—fulfilling freestanding criteria. To minimize mica residue, $La_{0.7}Sr_{0.3}MnO_3$ freestanding film preparation incorporates repeated blue-tape exfoliation: after depositing 60 nm $La_{0.7}Sr_{0.3}MnO_3$ on mica via PLD, room-temperature exfoliation with 3M tape yields freestanding films exhibiting large low-field magnetoresistance (LFMR) effects (Fig. 3b) [113]. Furthermore, Liu et al. [114] developed a single-step method for preparing transfer-free PZT thin films on mica, significantly simplifying the manufacturing process for flexible piezoelectric nanogenerators.

For epitaxial films that are strongly bonded to substrates, mechanical delamination is primarily achieved through stress-driven or interface-weakening mechanisms. The stress-driven approach involves either depositing high-stress control layers, such as Ni [115], Pt [116] or Al–Ag [50] stressors, onto the substrate or film surface to introduce controllable intrinsic stress by tuning deposition parameters, or inducing thermal stress arising from mismatches in the coefficients of thermal expansion during temperature variations. In contrast, the interface-weakening approach constructs van der Waals–bonded interfaces with reduced adhesion, thereby guiding crack propagation along predefined paths. Both methods require external mechanical forces (e.g., blade scoring, tape peeling, or fluid pressure) to initiate cracks, with crack penetration depth into the substrate governed by material fracture toughness and stress field distribution [68]. This process is stable for specific materials and widely applied in semiconductor film exfoliation.

Stress-driven exfoliation includes external and internal approaches. External exfoliation employs artificial stress sources: e.g., Dross *et al*. [50] induced parallel cracks via CTE mismatch between Ag/Al paste and silicon, obtaining 30–50 μm silicon membranes; for GaN, highly tensile-stressed Ni layers drive cracks along the 0001 plane when the stress-thickness product exceeds a critical value, enabling controlled release of 8–30 μm films [115] (Fig. 3c). More recently, it has been demonstrated that stress-engineered delamination can be achieved by designing stress gradients in sputtered Pt bilayers and further enhancing the stress through mechanical bending. In this approach, the controlled Ar pressure during sputtering creates distinct microstructures that lead to strong tensile stress gradients. When the strain energy release rate exceeds the interfacial fracture toughness, reliable delamination is enabled, allowing for damage-free dry transfer of high-temperature oxides [116]. Internal exfoliation exploits intrinsic stress: Sambri *et al*. [117] deposited supercritical-thickness (>46 nm) $LaAlO_3$ on $SrTiO_3$, where lattice mismatch and thermal stress initiated vertical cracks. Water molecule-induced interface weakening redirected cracks to propagate parallelly, forming self-curving micromembranes (Fig. 3d). This process can be guided by ion implantation or pre-patterned trenches to control dimensions (Fig. 3e, f) [118-121].

The interface energy modulation strategy enables precise exfoliation: when interfacial energy is lower than the surface energy of the film/substrate, cracks propagate cleanly along the interface without substrate damage. Liu *et al*. [55] developed a green, non-destructive van der Waals stripping (vdW stripping) technique. By sputtering Pt on mica to form a weak vdW interface, capillary forces exfoliate 2×2 cm$^2$ $Ba_{0.85}Ca_{0.15}Zr_{0.1}Ti_{0.9}O_3$ (BCZT) piezoelectric films (Fig. 3g). Atomic lift-off (ALO) technology weakens covalent bonds at perovskite/substrate interfaces (e.g., $PbTiO_3$/$SrTiO_3$ heterostructure) using Pb, combined with Cr/Ni stress layers to produce sub-10 nm ultrathin oxide films with atomically flat interfaces and superior functional properties (Fig. 3h) [122]. Theoretical calculations indicate that reduced interfacial charge transfer in Pb-containing materials underpins low interfacial energy. However, this technique is currently limited to Pb-based systems and requires enhanced mechanical stability and lead-free alternatives.

Successful mechanical exfoliation requires a core condition: the intrinsic work of adhesion at the target interface must be significantly lower than the critical fracture toughness of the film or substrate, ensuring cracks propagate preferentially along the interface without damaging the substrate material. This method avoids chemical processing, and is particularly suitable for materials with inherent cleavage planes, such as GaN and mica. The reusability of the substrate further reduces manufacturing costs, while the process enables the fabrication of high-performance functional films. However, its applicability is constrained by several factors, including high sensitivity to interfacial properties, strict requirements for uniform crack initiation and propagation, and the inevitable presence of residual substrate material on exfoliated films caused by uncontrollable cleavage depth, which hinders the universal realization of freestanding films. Fundamentally, this technique relies on stress concentration to induce directional cleavage along weak material interfaces, thereby achieving film separation.

**2.1.3 Remote Epitaxy van der Waals Lift-Off**

Remote epitaxy enables controllable thin-film growth and subsequent detachment by introducing a 2D material interlayer, such as transferred graphene, onto a single-crystal substrate. The process begins with transferring the 2D layer onto a pre-treated substrate (Fig. 4a), followed by epitaxial growth of the target film on top of it (Fig. 4c). Its functionality relies on two fundamental principles. In remote epitaxy, the lattice information of the substrate penetrates through the 2D material, directing the film to grow along the substrate orientation while the 2D layer suppresses strong chemical bonding, thereby preserving a van der Waals interface. In van der Waals lift-off, the weak interaction between the 2D layer and the substrate or film is exploited, enabling stress-free separation of the "2D layer/film" heterostructure without inducing damage (Fig. 4b). This process preserves film single crystallinity, and the resulting freestanding films can be directly utilized or transferred to arbitrary substrates for high-quality heterogeneous integration [111,123-126].

This technique's feasibility originates from the substrate polarity penetration mechanism [127]. Studies indicate that material ionicity directly governs electrostatic potential penetration capability: Low-ionicity materials (e.g., GaAs, ≈32%) penetrate only monolayer graphene, whereas high-ionicity materials (e.g., GaN, LiF) penetrate up to three layers, while covalent materials (e.g., Si, Ge) exhibit complete field shielding [128,129]. Density functional theory (DFT) further confirms that penetration depth correlates positively with ionicity and decays exponentially with distance [130], providing criteria for material selection.

Remote epitaxy evolved from the layered vdW epitaxy theory proposed by Koma [109], gaining rapid experimental validation: Chung *et al.* first demonstrated GaN epitaxy and lift-off on graphene/ZnO [131], proving the 2D interface's dual role as template and release layer. Subsequently, Kobayashi achieved full-wafer GaN detachment using h-BN as a release layer [132], whereas Kim *et al.* developed a recyclable graphene/SiC transfer process utilizing engineered strain layers (e.g., Ni thin films) for graphene exfoliation and transfer onto foreign substrates (Fig. 4a) with selective etching of excess layers to produce continuous monolayers [133].

Building on this principle, material systems have expanded continuously: The successful epitaxy of (001)-oriented $SrTiO_3$ on graphene-coated $SiO_2$/Si substrates enabled perovskite-silicon integration [134]; III-V compounds (GaAs/InP/GaP) were grown on graphene/GaAs [52] alongside sulfide materials (e.g., $HfS_2$) on h-BN/sapphire [72]; $CsPbBr_3$ films grown on graphene/NaCl exhibited dislocation densities reduced to 1/10 of direct epitaxial values with threefold-enhanced photoresponsivity [135]; Metal systems (e.g., Cu, GdPtSb alloys) and micro/nanostructures (e.g., ZnO microrods) [136-139] were further fabricated in combination with complex oxides including $BaTiO_3$, $CoFe_2O_4$, and $Y_3Fe_5O_{12}$ (Fig. 4c-e) [53,140]. These collective advances significantly broadened flexible electronics applications.

Critical to realizing these material systems is process optimization. First, metal stress layers (e.g., Ni) enabled wafer-scale lift-off and substrate recycling [52,136,137,140]. Notably, dry transfer increased III-V freestanding thin film fabrication success rates from <30% to >90% while avoiding wet-process oxidation risks. Moreover, in situ CVD-grown graphene on $Al_2O_3$/SiC substrates ensured III-nitride epitaxial reproducibility [130,141]. For growth techniques, MBE demonstrated high efficacy under ultra-high vacuum conditions, whereas metal organic CVD (MOCVD) required $N_2$ carrier gas to prevent graphene etching [129,142].

Recently, this technology has advanced toward systematic applications: bilayer graphene serves as the standard release layer for complex oxides. For instance, PMN-PT/$CoFe_2O_4$ heterostructures epitaxially grown on graphene/STO substrates have enhanced the magnetoelectric coupling coefficient by fivefold [51]. Diverse device-based applications further exemplify this progress, including graphene/GaN wireless surface acoustic wave (SAW) electronic skins (e-skins) (Fig. 4e) [140], h-BN/HfS$_2$ flexible photodetectors [71], and graphene/AlN deep-ultraviolet light-emitting diodes (LEDs) [141].

This technique achieves wafer-scale production of III-V compounds, oxides, and mixed-dimensional heterostructures through a "2D release layer + metal-stress lift-off" approach. Its success depends on synergistic optimization of three critical parameters: substrate polarity (ionicity >30%), interface cleanliness (dry transfer), and growth environment control (N$_2$ atmosphere in MOCVD), providing an efficient pathway for flexible electronics and heterogeneous integration. Despite significant advantages—including lattice mismatch tolerance and substrate recyclability—scaling challenges persist: wafer-scale graphene quality remains constrained by polycrystalline domains, while fundamental understanding of charge/strain mechanisms at mixed-dimensional interfaces requires breakthroughs in in situ characterization.

### 2.1.4 Alternative Physical Delamination Techniques

Besides conventional mechanical exfoliation and laser lift-off, ion beam etching and ultrasonic delamination have emerged as advanced physical delamination techniques with unique advantages for fabricating freestanding thin films. Unlike the shear-force-dominated mechanism of mechanical exfoliation and the photothermal ablation mechanism of laser lift-off, ion beam etching employs high-energy ion bombardment (e.g., Ar$^+$) to induce physical sputtering, achieving atomic-scale precision etching [28]. This technique, like laser processing, belongs to non-contact energy beam machining but induces less thermal damage to sensitive materials, making it particularly suitable for micro/nanofabrication of multilayer heterostructures (e.g., La$_{1-x}$Sr$_x$MnO$_3$/BaTiO$_3$/CeO$_2$/YSZ). Ultrasonic delamination fundamentally differs from the weak interfacial bond cleavage mechanism of vdW exfoliation: By leveraging microjets generated through cavitation effects, it overcomes interlayer forces in molecular crystals to achieve monolayer isolation of supramolecular complexes, which exhibits specific requirements for solvent polarity (e.g., acetonitrile) and interlayer energy differences [143].

Physical delamination methods (e.g., laser, mechanical, and remote vdW exfoliation) enable the fabrication of high-quality freestanding thin films, with inherent advantages including avoidance of chemical contamination, preservation of intrinsic properties, and recyclability of substrates. Nevertheless, significant limitations persist: process complexity and sensitivity, high equipment costs, restricted applicability (requiring specific interfaces/materials), and challenges in achieving large-area, ultrathin, defect-free delamination. Current research focuses on enhancing precision (sub-nanometer), efficiency (wafer-scale), and universality (strongly bonded materials). For films with ultra-strong interfacial bonds that resist physical delamination, chemical etching demonstrates unique potential through reaction selectivity.

## 2.2 Chemical Etching Techniques

Chemical etching represents one of the earliest techniques employed for fabricating freestanding films [4]. This approach leverages the selective etching action of specific chemical reactions to dissolve either the substrate material or a pre-designed sacrificial layer, thereby separating the film from its substrate. Based on the etching target, the technique is primarily categorized into two types: substrate etching and sacrificial layer etching. Compared to direct substrate etching, sacrificial layer etching has emerged as the most widely adopted technique due to its superior substrate versatility and enhanced process controllability. These two methods are detailed separately below.

### 2.2.1 Substrate Etching

This technique relies critically on etchant selectivity to dissolve the substrate material while preserving the thin film to remove interfacial constraints (Fig. 5a). For example, $SrRuO_3$ films are released by dissolving $SrTiO_3$ substrates in an $HF/HNO_3$ mixture, eliminating compressive strain of the film and restoring their magnetoelectric properties to bulk-like levels (Fig. 5e) [144]. Similarly, etching silicon substrates with HF solution releases $BiFeO_3$ films, relieving tensile strain to revert their crystal structure from monoclinic to rhombohedral phase. This transition enhances remnant polarization, demonstrating strain modulation of ferroelectric properties via polarization rotation mechanisms [145].

Dissolution of MgO substrates using 10% ammonium sulfate ($(NH_4)_2SO_4$) solution at 80 °C releases CFO single-crystalline films; subsequent transfer onto flexible substrates enables continuous tuning of magnetic properties via bending strain, while transfer onto silicon substrates eliminates magnetic anisotropy (Fig. 5a-c) [146]. Rapid etching of MgO with phosphoric acid at ~75 °C releases PZT nanoribbons which, upon transfer, form pre-strain-induced wavy buckling structures that enhance stretchability and piezoelectric response (Fig. 5d-e) [147,148]. Advanced implementations combine substrate pre-patterning and etched channel pore arrays: selective epitaxial growth of SrRuO$_3$ using amorphous MgO masks, followed by HF/HNO$_3$ etching of SrTiO$_3$, releases single-crystalline freestanding SrRuO$_3$ thin films (Fig. 5f). These exhibit ultrahigh flexibility and room-temperature heterogeneous bonding, enabling the first demonstration of oxygen partial pressure-modulated ferroelectric polarization switching in an oxide system, thus providing a novel platform for flexible electronics [149]. Additionally, Kelso *et al.* demonstrated that spin epitaxy combined with soluble templates (e.g., NaCl) offers an alternative route to obtain freestanding functional layers [11].

Released films typically require polymer supports (e.g., PDMS/PMMA) for transfer and annealing to repair interfacial defects. Substrate etching not only eliminates epitaxial strain to restore the intrinsic properties of films but also enhances mechanical adaptability through microstructural design of growth templates. For example, filling sealed nanopores in anodized aluminum oxide (AAO) templates followed by AAO dissolution creates nanopillar arrays, increasing the specific surface area of chitosan and enhancing cell adhesion efficiency and antibacterial properties of the films [150].

Substrate etching is a preferred strategy due to its relatively low equipment investment, high selectivity, and superior property tunability. It achieves selective substrate dissolution using chemical solvents (e.g., HF solution) under ambient pressure, featuring simple equipment and inherent compatibility with large-area substrates. This avoids the vacuum requirements of dry etching and the risk of lattice damage from physical bombardment [59,151,152]. Although this technique eliminates the need for sacrificial layer design, it faces challenges including high substrate cost, film fragility, and difficulties in large-area transfer. Its success critically depends on two factors: the high selectivity of the etching system, where the substrate-to-film dissolution rate ratio typically exceeds 100:1 to ensure film integrity, and the implementation of effective film protection strategies, such as polymer support layers like PMMA or PI tape, which enable crack-free transfer.

**2.2.2 Sacrificial Layer Etching**

The sacrificial layer etching technique achieves thin-film separation by introducing a material layer sensitive to specific chemical etchant between the substrate and functional film. This process initiates with depositing sacrificial layer materials (e.g., BaO [153,154], La$_{1-x}$Sr$_x$MnO$_3$ [54,155-157], SrRuO$_3$ [158], YBa$_2$Cu$_3$O$_{7-x}$ [159], Sr$_3$Al$_2$O$_6$ [34,74,75,160]) onto the substrate, followed by epitaxial growth or deposition of the target thin film on the sacrificial layer. The sample is then immersed in a selective etchant where precisely controlled etching parameters ensure complete dissolution of the sacrificial layer. Ultimately, the released freestanding thin film is collected for cleaning and post-processing procedures, including residue removal, annealing, and transfer.

This technique traces its origins to the intermediate-layer etching strategy proposed by Konagai *et al.* in 1978 [3], wherein freestanding GaAs films were successfully fabricated through selective etching of $Ga_{0.3}Al_{0.7}As$ layers using HF. Subsequent developments include water-soluble NaF/NaCl sacrificial layer system developed by Deutscher *et al.*[161], while Prodjosantoso [162], Vendamme [163], and Lazic [164] expanded material systems and processes via crystal structure engineering and organic-inorganic hybrid network design. The core mechanism of this technique lies in the differential chemical responsiveness to etchant among the sacrificial layer, functional thin film, and substrate. Material selection must satisfy four critical criteria: epitaxial compatibility, etching selectivity, high-temperature stability, and atomic-level controllability in fabrication.

Sacrificial layers are primarily categorized into water-soluble and water-insoluble types based on their dissolution characteristics. Water-soluble sacrificial materials (e.g., $Sr_3Al_2O_6$ [34,160,165], BaO [153,154], $SrVO_3$ [166]) can typically be removed using deionized water as the etchant. In contrast, water-insoluble sacrificial materials (e.g., $Ga_{0.3}Al_{0.7}As$ [3], AlAs [167] $TiO_2$ [168]) require specific chemical solvents for etching. As summarized in Table 2, particular solvents are necessary for certain materials: for instance, $YBa_2Cu_3O_{7-x}$ dissolves in dilute HCl [159]; $La_{1-x}Sr_xMnO_3$ etches in a KI + HCl mixed solution [54,155-157,169-171]; $SrCoO_{2.5}$ is soluble in acetic acid [172]; and $SrRuO_3$ requires sodium periodate ($NaIO_4$) solution for dissolution [158]. As illustrated in Fig. 6, these diversified material systems (encompassing binary compounds such as $TiO_2$/AlAs/BaO and perovskite-structured materials like $SrVO_3$/$La_{1-x}Sr_xMnO_3$/$YBa_2Cu_3O_{7-x}$ not only underpin the hetero-integration of ferroelectric/multiferroic films [54,158] and the development of flexible devices [146,155,156,173], but also enable novel processes such as intelligent deformation [169] and rapid non-destructive transfer [159].

The water-soluble sacrificial layer method, utilizing a mild etchant (deionized water), enables near-damage-free fabrication of freestanding thin films (Fig. 7d)—including multiferroic films as thin as a unit cell [174]. This technique has become mainstream, with alkaline-earth aluminates (e.g., $Sr_3Al_2O_6$) [175-188] and alkaline-earth oxides (e.g., SrO, BaO) [73,146,153,154] being representative materials. PLD dominates in growing complex sacrificial layers (e.g., $Sr_3Al_2O_6$, $Sr_4Al_2O_7$), while simpler oxides (e.g., BaO/SrO) are more suitable for direct molecular beam epitaxy (MBE) growth due to their structural simplicity [68].

The cubic-phase $Sr_3Al_2O_6$ sacrificial layer, owing to its water solubility and pseudo-cubic lattice parameter ($a^* \approx 3.96$ Å, corresponding to a cubic cell of $a = 4 \times a^* \approx 15.84$ Å) compatible with perovskite oxides [15,31,74,75,78,189], facilitates nontoxic etching and release of single-crystal oxide films and heterostructures, such as $BiMnO_3$ [31], $BiFeO_3$ [33,34,188], $La_{1-x}Sr_xMnO_3$ [15,187], $BaTiO_3$ [78,160,165,186,189], $VO_2$ [176], Pt/Co/Ni/Co heterostructure [30,190]. As representative examples of such released structures, Fig. 7e demonstrates SEM and ADF-STEM images of the superelastic freestanding $PbTiO_3$/$SrTiO_3$ superlattice tube [179] and the $SrRuO_3$/$BaTiO_3$/$SrRuO_3$ heterostructure film [181]. Beyond these structural prototypes, this release process unlocks emergent functional properties, as demonstrated by freestanding $BaTiO_3$ thin films which exhibit wrinkled morphology with zigzag patterns under ultralow strain (≈0.5%), inducing periodic polarization topologies (head-to-tail boundaries, vortex/antivortex domains) that enable new degrees of freedom for flexible devices (Fig. 7f) [78]. Extending beyond structural control, room-temperature deposition of amorphous $Sr_3Al_2O_6$ on $SrTiO_3$ substrates at room temperature generates an interfacial two-dimensional electron gas (2DEG). Dissolving $Sr_3Al_2O_6$ with water droplets removes interfacial oxygen vacancies, erasing the 2DEG. Repeated $Sr_3Al_2O_6$ deposition/dissolution enables programmable "writing/erasing" of the 2DEG. By modulating oxygen vacancy concentration, metal-insulator transitions (MITs) can be controlled, accompanied by a nonlinear-to-linear Hall effect shift, revealing dual-band transport (Fig. 7g) [191]. However, progressive interfacial stress accumulation during epitaxial growth of cubic $Sr_3Al_2O_6$ induces cracking, limiting film dimensions [74]. Additionally, its slow dissolution kinetics (hours for millimeter-scale separation) hinders wafer-scale applications [75,176].

The recent advent of $Sr_4Al_2O_7$ via epitaxial strain engineering marks a breakthrough in freestanding thin films fabrication [75,184,192,193]. As shown in Fig. 7a, the low-symmetry crystal structure of $Sr_4Al_2O_7$ (transformable from orthorhombic to tetragonal under biaxial strain) and widely tunable pseudo-cubic in-plane lattice parameters (3.808–4.035 Å, DFT-calculated $a^* = b^* = (a^2 + b^2)^{1/2}/4 \approx 3.896$ Å, out-of-plane $c^* = c/6 \approx 4.288$ Å) significantly enhance perovskite lattice matching. This enables millimeter-scale crack-free freestanding membranes of nonferroelectric oxides and broadens compatibility with diverse oxide systems (Fig. 7c) [75,174], addressing post-release performance degradation bottlenecks. The discontinuous Al-O framework and Sr-O-rich networks of $Sr_4Al_2O_7$ (Fig. 7a) accelerate water dissolution by 10× compared to $Sr_3Al_2O_6$ [184]. Its low symmetry also suppresses interfacial strain, while broad epitaxial growth windows (Fig. 7b) provide processing flexibility. These synergistic advantages position $Sr_4Al_2O_7$ as the most scalable solution for freestanding thin film production, laying the foundation for flexible spintronics.

Sacrificial layer advancements directly enhance functional film properties through two universal mechanisms: strain relaxation and interface optimization. Specifically, strain relaxation induces an MA-to-R phase transition in freestanding $BiFeO_3$, amplifying Fe-ion displacement and polarization to boost photovoltaic response by ≈200% in 50-nm films—though this effect diminishes with increasing thickness while maintaining single-domain states [33]. Concurrently, $Sr_3Al_2O_6$-based freestanding ferroelectric membranes achieve flexibility, with a $BaTiO_3/SrTiO_3$ superlattice design (11×$SrTiO_3$ layers) enhancing polarization by 95.92% through strain preservation without interfacial interdiffusion, thereby stabilizing ferroelectricity [27].

Furthermore, substrate removal universally enhances magnetic and electrical properties (Fig. 7h). Lattice relaxation elevates Curie temperatures ($T_C$) in perovskite manganite films through optimized Mn-O-Mn bonding, with distinct enhancement effects observed across material systems: $La_{0.67}Sr_{0.33}MnO_3$ exhibits a moderate $\Delta T_C$ of 10–15 K [187], while $La_{0.7}Ca_{0.3}MnO_3$ and $(La_{2/3}Pr_{1/3})_{5/8}Ca_{3/8}MnO_3$ (LPCMO) demonstrate significantly larger gains of 137 K and 29 K, respectively, attributable to dead-layer elimination [15,185]. Concomitantly, this relaxation process suppresses electronic phase separation in LPCMO, thereby reinforcing ferromagnetic order. Notably, the successful fabrication of freestanding superconducting $La_{0.8}Sr_{0.2}NiO_2$ films, exhibiting a superconducting transition temperature of 10.6 K, was enabled by two key innovations. The first involved suppressing precursor defects through a $TiO_2$-terminated $SrTiO_3$ buffer layer combined with LaO interfacial passivation. The second was a rapid, near-damage-free transfer completed in under two minutes, facilitated by $Sr_4Al_2O_7$ sacrificial layers that preserved the metastable infinite-layer structure [193].

Beyond conventional chemical etching, emerging pathways to freestanding architectures are based on the thermal decomposition of sacrificial organic frameworks. In these approaches, functional precursors are first constrained and shaped by surface tension within 3D-printed polymer meshes. During subsequent sintering, pyrolysis of the organic scaffold chemically removes the supporting structure, thereby releasing the freestanding ceramic component. This mechanism is exemplified by the surface tension-assisted two-step (STATS) process for fabricating programmable cellular ceramics [194], and the stress-eliminated liquid-phase fabrication (SELF) method, which produces crack-free thick films (1–100 μm)—the latter achieving a notable piezoelectric coefficient ($d_{33} \approx$ 229 pC N$^{-1}$) in freestanding PZT [195]. It should be noted that such pyrolysis-based methods may be limited in producing epitaxial films with precise crystallographic orientation, and the attainable dimensions of individual units are constrained by capillary stability. Nonetheless, their unique capability to directly fabricate complex 3D architectures and high-performance thick films offers an attractive strategy for developing flexible and conformal electronics.

In summary, sacrificial layer etching offers a broadly compatible and highly controllable route for releasing freestanding thin films, with demonstrated advantages in material versatility, interface cleanliness, and 3D structuring potential. Co-optimizing sacrificial-layer composition, interface quality, and strain evolution further enables high-performance functional membranes. However, persistent issues such as release adhesion, stress-induced deformation, and chemical residues remain intrinsic challenges within this etching-based pathway, motivating continued development of smarter sacrificial layers and cleaner release chemistries for large-area and heterogeneous integration.

## 2.3 Core Physical Principles and Fabrication Mechanisms

Apart from the distinctions among individual techniques, all freestanding thin-film fabrication routes are fundamentally unified by a common physical essence: achieving controlled separation by overcoming the interfacial binding energy between the film and the substrate through energy regulation. The common principle underpinning all these technical pathways is the precise manipulation of interfacial energy, strain, and stress. The initiation of delamination can be fundamentally attributed to an energy competition. Whether it involves the laser-induced thermal stress in laser lift-off, stress concentration in mechanical exfoliation, or selective dissolution in chemical etching, the success of each method hinges on a critical condition: the introduced external or internal energy (mechanical, thermal, or chemical) must be sufficient to disrupt the energy balance at the interface, driving the total system energy towards separation. For instance, mechanical exfoliation requires the work of adhesion at the interface to be significantly lower than the critical fracture toughness of either the film or the substrate itself, ensuring crack propagation along the interface rather than through the bulk of the film.

Furthermore, the key to achieving clean delamination lies in interface design. The fundamental distinction among different technical pathways resides in their approach to treating and managing this interface. They form a technological spectrum ranging from strategies dealing with "strong interfaces - forced separation" (e.g., some forms of laser lift-off and mechanical spalling) to those utilizing "weak interfaces - gentle separation" (e.g., remote epitaxy and atomic lift-off). The latter approaches pre-construct a weakly bonded interface by introducing van der Waals gaps or chemical modifications, drastically reducing the energy required for separation and thereby enabling the production of damage-free, ultrathin films, representing a leading direction in technological advancement.

Moreover, the ultimate objective of delamination is to liberate the intrinsic properties of the material. Once freed from substrate clamping, strain relaxation emerges as a universal and direct consequence. This represents not merely a geometric alteration but a functional emancipation: it enables polarization rotation in ferroelectric materials, modulation of magnetic anisotropy, and the exploration of quantum states in strongly correlated electronic systems. In summary, delamination technology transcends its role as a mere fabrication step, establishing itself as a potent tool for strain engineering and property control.

## 3. Transfer and Integration Techniques

The successful release of thin films from their substrates is a critical prerequisite for achieving a freestanding state. However, release is merely the initial step; the core challenge lies in the damage-free and precise transfer of the detached, free-floating, and highly fragile films onto target substrates or device architectures. This transfer process is decisive for their successful application in subsequent research or functional devices.

### 3.1 Wet Transfer Technique

Currently, two primary process routes have been developed for the heterogeneous integration of freestanding functional films via wet transfer, which exploit the combined use of sacrificial layers and polymer carriers. The first is the "polymer carrier–sacrificial layer" route, while the second is the "self-supporting" route. The former employs temporary carriers (e.g., PMMA, PS, or photoresist) combined with physical/chemical fabrication of freestanding films, enabling large-area transfer with minimal cracking. The latter entirely omits polymer carriers, relying on the film's buoyancy (or utilizing the original substrate as temporary support) to prevent fracture. While this approach simplifies the process, it demands higher intrinsic film toughness [146].

The key to wet transfer of freestanding films lies in the pre-fixation of temporary carriers and precise control of the etching process. As shown in Fig. 8a, prior to sacrificial layer etching, temporary carriers (e.g., PMMA, EVA) are firmly attached to the film surface via spin-coating or tight bonding [55,159]. The composite structure is then immersed in a selective etchant to dissolve the sacrificial layer. After etching, the film-carrier complex is released and floats on the liquid surface. Following residue removal through rinsing, the freestanding film is aligned and scooped onto the target substrate using tweezers or positioning tools. Van der Waals forces between the film and the new substrate can be enhanced by thermal treatment, after which temporary carriers (e.g., PMMA) are ultimately removed in acetone solution.

Wet transfer has been successfully applied to heterogeneous integration of diverse functional oxide films. For instance: $La_{1-x}Sr_xMnO_3$ [159] and LPCMO [185] (Fig. 8c) manganite films transferred to silicon wafers using polymer carriers (PMMA, PPC) and water-soluble sacrificial layers ($YBa_2Cu_3O_{7-x}$, $Sr_3Al_2O_6$) exhibit effective strain relaxation and optimized magnetoelectric properties; PZT ferroelectric films [54] and $NaNbO_3$ antiferroelectric films [171] transferred onto silicon-on-insulator (SOI) and $SiN_x$ substrates, respectively, via analogous methods achieve high-quality interfaces and significant electrical property modulation; $BiFeO_3$ films are transferred using photoresist and $Sr_3Al_2O_6$ ( sacrificial layers; notably, $BaTiO_3$ films transferred via self-supporting flotation onto $SiO_2$/Si substrates retain fundamental ferroelectricity despite crack formation [33].

In summary, the synergy between the "soluble sacrificial layer" and the "temporary carrier" in wet transfer enables large-area, low-damage integration of ferromagnetic, ferroelectric, antiferroelectric, and conductive oxides onto silicon-based and flexible substrates, significantly advancing functional oxides for heterostructure devices and flexible electronics. The liquid-mediated environment effectively mitigates electrostatic adhesion and repulsion, while polymer carriers provide essential mechanical support, stress buffering, and cleaner transfer interfaces by dissolving sacrificial-layer residues. Nevertheless, compared with certain dry-transfer methods, wet transfer still faces several challenges. These include potential chemical incompatibility, such as film degradation from etchant or solvent residues; limited precision in interfacial adhesion control, which strongly depends on substrate surface treatment; film wrinkling or cracking driven by capillary forces and surface tension; and both inefficiency and contamination risks arising from its multi-step processing, including spin-coating and peeling.

**3.2 Dry Transfer Technique**

Dry transfer utilizes van der Waals forces/adhesion between a solid medium (e.g., PDMS, PI, other thermal release tapes) and freestanding films to achieve solvent-free relocation, thereby avoiding structural damage induced by solvent capillary forces [79,116,122,156,158,182,187]. This technique has become a central approach for manipulating freestanding films, particularly excelling in the transfer of functional perovskite films released from sacrificial layers [74,75]. The process begins with gently contacting the film surface using a PDMS stamp at low speed and a slight tilt to remove air bubbles. The sacrificial layer is then chemically etched to yield an intact freestanding film, and the transfer is finalized using a micrometer-aligned stage (Fig. 8b) [15,177].

The carrier-sacrificial layer synergy strategy enables universal transfer through differential design, with validated performance across ferroelectric, ferromagnetic, and superconducting systems. PDMS's low surface energy facilitates rapid peeling of millimeter-scale films [34,74]: for instance, freestanding $BiFeO_3$ achieved giant polarization in the 2D limit via lift-off and transfer [34], while $La_{0.7}Ca_{0.3}MnO_3$ attained >8% uniaxial strain using a $SrCa_2Al_2O_6$ sacrificial layer and PI carrier, triggering an antiferromagnetic insulating phase beyond critical strain [77]. Furthermore, dry transfer of $La_{1-x}Sr_xMnO_3$ single-crystal films with water-soluble $Sr_3Al_2O_6$ sacrificial layers significantly enhanced $T_C$ [187]; films transferred onto PMN-PT substrates exhibited electrically controlled magnetism, where strain from the ferroelectric substrate enabled electrical control of magnetic anisotropy [158].

Dry transfer technology utilizes vdW force-mediated solid-solid interfacial interactions to circumvent material degradation induced by solvent infiltration, providing a non-destructive transfer pathway for water/oxygen-sensitive systems (e.g., nickel-based superconducting films).

Freestanding films obtained via physical exfoliation can all be transferred through this method, with rigid templates including partially exfoliated substrates (e.g., mica [112]), stress-inducing metal layers (e.g., Ni [115], Pt [116]), or polymers such as PDMS. Freestanding films prepared by chemical etching can also achieve rapid separation and transfer without polymer assistance through metal stress assistance (e.g., Cr/Au bilayers) [167] (Fig. 8e). The optimal transfer strategy should be determined by comprehensive consideration of material properties, target substrates, required transfer precision, and post-transfer application requirements.

Current dry transfer techniques face several critical challenges. Bubbles formed at the transfer interface due to environmental adsorbates can induce localized strain, leading to performance degradation. Mismatched adhesive forces restrict the transferable dimensions of freestanding films and may cause tearing, while strain relaxation during the process can result in uncontrollable wrinkling and crack propagation.

To address these bottlenecks, cross-scale synergistic strategies can be implemented: For mechanical reinforcement, composite support layers are employed to enhance film fracture toughness, while progressive stress release enabled through optimized transfer angles combined with thermally releasable tapes [157]; for interface management, transfer integration is conducted in ultraclean environments to suppress contamination, with interfacial bubbles eliminated by in situ annealing; for strain optimization, pre-strain compensation is implemented via lattice constant matching between substrates and films [175]. Strain-adjustable oxide templates (e.g. $Sr_4Al_2O_7$) can actively regulate stress relaxation through tunable lattice parameters and coherent strain interfaces, thereby suppressing cracks while preserving controlled wrinkles to achieve high-quality freestanding thin films [75]; for defect control, sacrificial layer dissolution rates can be modulated by thickness adjustment to mitigate film defects and cracks [189].

Future work could employ probe-based micro-manipulation systems to achieve stress-free transfer of micron-scale freestanding films [120], with non-contact precision handling enabling heterostructure integration as demonstrated in Fig. 8f.

## 3.3 Rigid-Flex Dual-Protection Transfer

The rigid-flex dual-protection transfer technique is a method for fabricating functional thin-film heterostructures based on sacrificial layer release. Its core principle lies in the synergistic effect between a rigid frame (exemplified by PMMA) for deformation suppression and a flexible polymer (e.g., PET/PDMS) for encapsulation protection, enabling the intact, damage-free transfer of millimeter-sized single-crystalline thin films and the preservation of their physical properties. As illustrated in Fig. 8d, the key process steps are as follows: first, the sacrificial layer and the target functional film are sequentially deposited onto a single-crystalline substrate; subsequently, PMMA is spin-coated onto the film surface as a temporary rigid support layer; finally, adhering an elastomeric stamp (e.g., PDMS) to the top, directly pressing it onto the PMMA/film composite structure, and immersing it into an etchant to dissolve the sacrificial layer; the released freestanding film is then transferred to a new substrate, and polymers like PMMA are ultimately removed using solvents such as acetone.

This technique offers an innovative solution to the challenges of strain control and interface protection during the transfer of functional thin films, wherein PET is employed for stress buffering [67]. Concurrently, the rigid PMMA layer ensures interface integrity through full-coverage encapsulation: In $BaSnO_3$ film transfer [196], PMMA isolates the film from water/oxygen corrosion, preserving an atomically smooth surface; for $Nd_{0.8}Sr_{0.2}NiO_2$ superconducting film transfer, PMMA encapsulation enables the film to retain a superconducting transition temperature of 17 K post-transfer [197].

Apart from transfer method, the choice of target substrate critically determines the final performance of freestanding films [15,173,187]. Transferring films onto flexible polymer substrates such as PET provides substantial tunability of mechanical strain, whereas transfer onto silicon wafers facilitates heterojunction formation and leverages compatibility with standard semiconductor processing. Transfer onto single-crystalline oxide substrates, such as $SrTiO_3$ or mica, enables precise modulation of magnetoelectric properties—including magnetic anisotropy, Curie temperature, and magnetoresistance—through controlled crystallographic orientation and interface coupling. Collectively, the substrate's mechanical properties, lattice matching, and interfacial chemical characteristics govern the film's strain state, charge transport behavior, and magnetic ordering, offering essential design parameters for multifunctional devices.

The transfer techniques discussed above represent mature and precise methods for heterogeneous integration. Meanwhile, innovative approaches such as flame-treated spray (FTS) coating allow for rapid, energy-efficient, and cost-effective conformal fabrication of piezoceramic films on free-form substrates [198]. Additionally, strategies that circumvent conventional transfer steps, such as the STATS [194] and SELF [195] methods for in-situ pyrolysis of sacrificial polymer frameworks, offer another complementary route toward scalable device integration. Beyond these integration-focused methods, the active self-assembly of functional biomolecular films—such as piezoelectric *β*-glycine layers guided by synergistic nanoconfinement and in-situ electric fields—enables the direct, bottom-up construction of freestanding crystalline films without relying on external transfer or sacrificial layer removal [199]. These approaches collectively broaden the technological toolkit for freestanding electronics.

**3.4 Transfer-Enabled Planar-View TEM Technique**

The advent of freestanding thin-film fabrication and transfer techniques has not only established a cross-material strain engineering platform for flexible electronics and quantum devices but also provided a novel and powerful technical pathway for direct planar-view structural characterization of thin films. Compared to traditional methods relying on focused ion beam (FIB) etching for sample preparation—which are typically costly, technically demanding and time-consuming—the emerging approach based on sacrificial layer etching combined with dry/wet transfer techniques demonstrates significant advantages.

The core of this methodology (schematic shown in Fig. 9a) involves the controlled release of target films from substrates via selective etching of pre-designed sacrificial layers, yielding large-area, stress-free freestanding films. These films are then precisely transferred onto transmission electron microscopy (TEM)-compatible microgrids or support meshes (Fig. 9b) using transfer techniques. This strategy enables direct TEM observation of film planes, circumventing the viewing-angle constraints of conventional cross-sectional analysis. It allows direct probing of planar-view microstructures (e.g., grains, defects, domains, and strain fields), assessment of homogeneity, and supports multimodal analyses including selected-area electron diffraction, high-resolution TEM, scanning TEM, electron energy-loss spectroscopy, and energy-dispersive X-ray spectroscopy.

This technique has been extensively adopted in cutting-edge materials research. Numerous studies have successfully utilized it to elucidate planar crystallinity, microstructural evolution, defect engineering, and atomic-scale interfacial information in functional materials such as complex oxide films and semiconductor thin films (Fig. 9c), providing critical experimental evidence for understanding the structure-property relationships in these systems [154,157,169-171,173].

## 3.5 Controlled Twistronics Integration in Freestanding Films

Recent advances in freestanding thin-film fabrication and transfer techniques have freed complex oxides from substrate-imposed epitaxial constraints, enabling unprecedented control over geometric stacking parameters and offering a new materials platform for twistronics. Inspired by the emergence of correlated insulating states, flat bands, and unconventional superconductivity in twisted van der Waals 2D materials [200-202], researchers have begun to extend twist-angle engineering into freestanding oxide membranes. Nevertheless, the implementation of twistronics in thin films differs fundamentally from that in 2D material systems due to the strong covalent/ionic interlayer bonding, strain relaxation pathways, and interface reconstruction unique to complex oxides.

### 3.5.1 Twist-Angle Control and Verification

Precise twist-angle control ($\alpha$) in freestanding films is typically established during stacking, using PMMA-assisted wet transfer, PDMS dry pickup, or vdW assembly (Fig. 10a–d). Alignment begins at the macroscopic level by cutting the thin films along known crystallographic edges, followed by optical microscopy–based azimuthal pre-alignment. The final $\alpha$ is accurately determined through nanoscale crystallographic verification, most commonly by X-ray diffraction $\varphi$-scans [26].

In 2D materials, the weak interlayer coupling allows the twist angle to be adjusted even after stacking, either through Atomic Force Microscope tip [203-205] manipulation or by thermally induced rotation [206-208]. In contrast, the strong covalent and ionic bonding in oxide films causes the layers to lock immediately upon contact, so the twist angle cannot be modified afterward [209]. This fundamental distinction has profound implications for moiré formation and emergent physics.

### 3.5.2 Distinct Moiré Engineering in Oxides

In vdW materials, moiré patterns emerge continuously with a geometric period $\lambda = a/(2 \sin (\alpha/2))$ (where a is the lattice constant) [210,211]. In contrast, thin films exhibit strain-mediated moiré physics strongly influenced by interfacial energies. At small $\alpha$ (<5°), twist-induced lattice mismatch is often accommodated not by long-range moiré fringes but through networks of screw dislocations and local interface reconstruction. Well-defined moiré patterns typically appear only above several degrees, and even then their period deviates from the ideal geometric model due to strain relaxation and oxygen-octahedra tilting. These interfacial structures—either dislocation grids or true moiré superlattices—are experimentally verified by transferring the bilayers onto TEM grids for HRTEM imaging [212,213]. This non-vdW behavior highlights the central role of strong interlayer coupling in freestanding thin film twistronics.

**3.5.3 Emergent Opportunities Beyond 2D vdW Twistronics**

The combination of substrate decoupling, precise twist control, and strong interlayer coupling in freestanding thin films creates unique opportunities inaccessible in both conventional epitaxy and vdW materials:

(1) Tunable Correlated States

Long-range moiré potentials reshape electronic structures, enabling flat-band formation, Dirac point reconstruction, and correlation-enhanced phases [209]. Strong coupling additionally permits twist-controlled modification of orbital hybridization and interfacial charge transfer—key parameters for engineering correlated ground states [214-216]. Twisted $La_{0.8}Sr_{0.2}CoO_3$ bilayers, for example, show twist-angle–dependent Curie temperatures, where weakened O–Co hybridization suppresses magnetism but gradually recovers with larger $\alpha$ [212].

(2) Topological and Spin–Orbit Phenomena

Twist-induced symmetry breaking enables chiral spin textures, polar vortices, and topological transitions—effects inaccessible in vdW systems due to their weaker structural coupling [26,209].

(3) Moiré Excitons, Orbital Magnetism, Quantum Geometric Responses

Moiré potentials in twisted oxide heterojunctions (e.g., $SrTiO_3/Ce_{0.8}Gd_{0.2}O_{1.9}$ [26]) generate correlated excitonic states, orbital magnetism, and Berry curvature–related quantum geometric responses, offering avenues for high-efficiency energy and neuromorphic devices.

(4) Stacking-Configuration–Dependent Interfacial Physics

Differences between AA- and AB- type stacking modulate interfacial dipole orientations and band alignments, directly influencing valley polarization and spin transport, thereby enabling new device concepts in oxide valleytronics and spintronics [26,213].

Freestanding films can also serve as crystalline templates for oriented growth of functional layers such as $BiFeO_3$ or $La_{1-x}Sr_xMnO_3$, enabling homostructure construction on arbitrary substrates (Fig.

10e–h). When the freestanding film and receiving substrate share matching lattice constants—e.g., (110)-SrTiO$_3$ on (110)-SrTiO$_3$—a preset twist angle is typically required to achieve coherent stacking [217]. In mismatched systems—e.g., (001)-SrTiO$_3$ on (001)-LaAlO$_3$—strain engineering is possible even without intentional twist due to spontaneous relaxation [218].

Twist-angle integration in freestanding oxide films not only circumvents the lattice-matching limitations of traditional epitaxy but also introduces a highly programmable geometric degree of freedom for designing correlated, topological, and functional quantum materials. Looking forward, merging high-precision twist control with *in situ* optical, electronic, and magnetic probes may enable dynamically reconfigurable moiré quantum devices and push oxide twistronics towards practical applications in quantum simulation, neuromorphic computing, and topological electronics.

## 4. Properties and Applications of Freestanding Thin Films

### 4.1 Core Characteristics and Performance of Freestanding Thin Films

#### 4.1.1 Structural Integrity Modulation

Thickness and lattice matching synergistically govern structural integrity. Ultrathin freestanding films fabricated via sacrificial layers achieve >10% recoverable superelasticity through strain gradient design, exemplified by continuous a/c domain rotation in BaTiO$_3$ [160]. And millimeter-scale crack-free transfer necessitates lattice matching between sacrificial and functional layers [75]. The compatibility of these parameters critically modulates interfacial stress distribution, thereby suppressing cracks and ensuring film integrity. This synergistic control constitutes the fundamental requirement for large-scale crack-free freestanding thin films (Fig. 11) [1,74,189].

#### 4.1.2 Interface Decoupling Effect

Upon removal of substrate clamping, materials can exhibit extraordinary properties in the low-dimensional limit that are inaccessible through conventional epitaxy. This behavior arises from a synergistic combination of three key effects. First, strain relief fully eliminates compressive or tensile strains caused by thermal expansion mismatch or lattice misfit, restoring the intrinsic lattice configuration, as exemplified by the recovery of the rhombohedral phase in BiFeO$_3$, which enhances polarization [33]. Second, interface reconstruction improves the chemical and electronic quality at the substrate-film interface by removing defects such as oxygen vacancies and dangling bonds, enabling pristine van der Waals contacts; for instance, the dissolution of a $Sr_3Al_2O_6$ sacrificial layer eliminates the two-dimensional electron gas at the $SrTiO_3$ interface [191]. Third, dimensional confinement introduces dominant low-dimensional quantum effects, triggering unique quantum behaviors that cannot be realized in bulk materials, as demonstrated by the preservation of superconductivity with modified transition temperatures in freestanding $La_{0.8}Sr_{0.2}NiO_2$ and $Nd_{0.8}Sr_{0.2}NiO_2$ thin films [193,197]. Collectively, these effects establish freestanding thin films as a transformative platform for exploring correlated electronic states, topological quantum phenomena, and low-dimensional quantum effects.

**4.1.3 Extreme Mechanical Properties**

Breakthroughs in the extreme mechanical properties of freestanding thin films represent a fundamental paradigm shift in strain engineering. Moving beyond incremental improvements within the conventional framework of substrate constraint and uniform strain, advanced fabrication techniques for freestanding thin films allow the exploration and utilization of emergent physical effects such as strong strain gradients and flexoelectric or flexomagnetic phenomena. These advances achieve mechanical performance approaching theoretical limits at nano-to-micro scales, including ultralarge strains, exceptionally strong gradients, and abrupt atomic-level transitions, thereby opening new dimensions for material manipulation and device applications. The key breakthroughs can be categorized into four main aspects:

(1) Superelasticity and Ultra-large Strain via Ferroelectric Domain Switching and Dipole Rotation

As shown in Fig. 12a,c,d, BTO thin films achieve ~10% recoverable strain (approaching the 12% theoretical limit for bulk single crystals) through high-strain-gradient designs in $La_{0.7}Sr_{0.3}MnO_3$ electrode structures or multilayer scroll configurations. This superelasticity far exceeds the fracture threshold of conventional epitaxial films, while piezoelectric responses are significantly enhanced near the neutral strain plane due to flexoelectric effects [160,165]. Similarly, $La_{1-x}Sr_xMnO_3$/$BaTiO_3$ nano-spring heterostructures exhibit >500% elongation – 50 times greater than the plastic deformation capacity of copper nanowires – enabled by coherent interfacial strain transfer (Fig. 12b,e,f) [178]. This behavior originates from continuous dipole reorientation buffering, as opposed to dislocation slip in conventional materials.

(2) Synergistic Enhancement of Strength and Flexibility

As depicted in Fig. 12h, BiFeO$_3$ thin films at the 2D limit (thickness <10 nm) exhibit a yield strength of 7.5 GPa (8× higher than bulk) while sustaining bending radii down to 500 nm without fracture. This resolves the classical "strength-flexibility trade-off" dilemma and simultaneously achieves strong polarization (140 μC/cm$^2$, 40% above bulk) in substrate-free unit-cell-thick films [34]. This study confirms that such 2D phase transitions occur without requiring epitaxial strain, overturning conventional wisdom on strain-mediated ferroelectric polarization. The realization of ultra-large uniaxial strain (>6%) stems from the elimination of substrate pinning in freestanding films, enabling reconstruction of ferroelectric order parameters in PbTiO$_3$ (Fig. 12g) [76].

(3) Strain Programmability

Preconfigured buckling induced by sacrificial layer thickness modulation enables wavy morphologies in PZT nanoribbons, enhancing stretchability by 300% and piezoelectric response by 220% [147]. This controllable microstructure design establishes a new paradigm for biomimetic devices (e.g., artificial muscles).

(4) Strain-Gradient-Induced Novel Properties

Strong strain gradients in micron-scale corrugated GdPtSb films break symmetry via the flexomagnetic effect, enabling room-temperature transition from antiferromagnetic to ferromagnetic states with efficiency surpassing homogeneous strain [139]. Meanwhile, atomic-level abrupt strain layers (e.g., a 4-unit-cell transition zone in manganites) exhibit lattice spacing jumps from 379 pm to 388 pm, demonstrating strain boundary tailoring of electronic states [218].

Collectively, these breakthroughs establish a core technological framework for extreme mechanical regulation, rooted in the synergy of flexoelectric effects, surface electric field driving, strain-gradient-mediated magnetoelectric coupling, and local lattice relaxation. Future devices – from mechanically/electrically dual-mode FeFETs [25] to flexible unit-cell-thick sensors [34] – reveal a new paradigm connecting macroscopic mechanics with quantum properties.

**4.2 Device Application**

As illustrated in Fig. 13, this section systematically presents breakthrough device validations and applications of freestanding thin-film transfer and integration technology across five key domains:

high-density memory devices, optoelectronic devices, advanced energy storage/conversion, flexible electronics/biomedical applications, and quantum material state exploration. Collectively, these examples demonstrate the technology's robust potential as a versatile platform for multidisciplinary applications.

### 4.2.1 High-Density Memory Devices

The fabrication and transfer of freestanding thin films significantly enhance the integration density and performance of high-density memory devices. Transferred $PbTiO_3/SrTiO_3$ bilayers prepared via $Sr_3Al_2O_6$ sacrificial layers achieve polar nanodomain arrays with storage densities exceeding 200 Gbit/in$^2$ [219]. Similarly transferred $BaTiO_3$ thin films exhibit domain-wall memory with switching ratios up to $10^3$ [220]. The interlayer laser lift-off (ILLO) technique enables successful transfer of 1 kbit 1S-1R RRAM arrays from glass substrates to flexible platforms [29]. Furthermore, ultrathin (≈ 4 nm) magnetic Pt/Co/Ni/Co heterostructures transferred via this method demonstrate strong perpendicular magnetic anisotropy and efficient domain-wall motion [190]. These advances confirm that the technology enhances storage density, device performance, and substrate compatibility through high-quality heterogeneous integration, providing a reliable pathway for next-generation high-density memory.

### 4.2.2 Optoelectronic Devices

Damage-free transfer techniques are critical for flexible optoelectronics. Kim *et al.* [70] achieved selective transfer of GaN films using 266-nm nanosecond-laser-induced thermal decomposition, albeit with interfacial roughening due to thermal damage. In contrast, Sun *et al.* [97] employed 355-nm picosecond-laser-triggered cold ablation, forming ultrahigh-aspect-ratio interfaces that significantly reduced GaN surface roughness and residual stress. Both techniques enabled flexible LED transfer: Kim's approach permitted single-LED repair, while Sun's method transferred intact devices to tantalum foils, maintaining stable 450-nm electroluminescence under bending. This highlights the pivotal role of low-damage transfer in preserving optoelectronic performance.

### 4.2.3 Advanced Energy Storage and Conversion

Freestanding film transfer enhances energy and catalytic devices by enabling strain engineering and 3D structuring. Park *et al.* [100] transferred large-area PZT films to flexible PET via laser lift-off, fabricating piezoelectric devices that generate 200 V voltage and 150 μA·cm$^{-2}$ current density under microstrain. Wang *et al.* [181] released SRO films using water-soluble sacrificial layers and rolled them into multilayer architectures, expanding electrochemical active area by 18.5× while reducing oxygen evolution overpotential by 74–78% in acidic/alkaline environments through strain-induced $Ru^{4+}$ spin-state transitions. These studies establish new paradigms for efficient energy conversion and catalysis via precise control of structural degrees of freedom.

**4.2.4 Flexible Electronics and Biomedical Applications**

This technology serves as a crucial enabler for biomedical flexible electronics. Laser-lift-off PZT films with high flexibility and piezoelectricity drive curved artificial cochlear hair cell sensors (iPANS) GaN micro-LED arrays transferred via laser lift-off were implanted into murine cortex for optogenetic stimulation [35]. BCZT films transferred by aqueous lift-off and integrated into PDMS resonators achieved high-sensitivity SARS-CoV-2 protein detection (frequency shift sensitivity: 100 kHz) [55]. Furthermore, the potential of biological materials is being increasingly unlocked. For instance, ultrathin biopiezoelectric submucosa films fabricated via van der Waals exfoliation have been demonstrated for human motion monitoring [221]. In a notable advance, actively self-assembled *β*-glycine films exhibit a combination of natural biocompatibility and robust piezoelectric output, showing significant promise for implantable biomedical electronics [199]. Key to these applications is efficient device integration on flexible bio-platforms through interfacial weakening, crystal structure optimization, and biocompatible encapsulation.

In addition, the emergence of freestanding MXene membranes provides a new materials platform for constructing multifunctional flexible systems. These membranes possess inherent flexibility, and their electrochemical activity and ordered nanostructures can be simultaneously exploited for flexible actuators [222] and high-sensitivity sensors [223]. More intriguingly, the same flexible architecture demonstrates breakthrough performance in gas separation membranes [224], indicating its significant potential for future applications in intelligent wearable systems and even flexible, miniaturized separation units.

**4.2.5 Quantum State Exploration**

Freestanding film technology enables *in situ* strain and carrier concentration control, serving as a vital platform for probing quantum phenomena in the 2D limit. Millimeter-scale $La_{0.8}Sr_{0.2}NiO_2$ films fabricated by MBE and sacrificial-layer exfoliation [193,197] exhibit superconducting transition temperatures ($T_S \approx 10.6$ K) comparable to bulk materials, confirming superconductivity robustness under dimensional reduction. Cryogenically transferred Bi-2212 monolayers [43] show unexpectedly high $T_S \approx 90$ K, challenging theoretical predictions, while Nb$_3$Cl$_8$ monolayers [44] exhibit strong correlated Mott gaps (bandwidth $\approx 0.3$ eV, $U \approx 1.2$ eV) via ARPES and computational verification. These approaches provide powerful tools for *in situ* manipulation of quantum states including superconductivity and Mott insulation.

### 4.3 From Property Enhancement to Functional Devices

It is particularly crucial to recognize the profound intrinsic connection between the fabrication pathways and the resulting functional properties. The liberation from substrates and the acquired freedom for strain modulation create essential conditions for restoring and optimizing the material's intrinsic state. Specifically, strain relaxation releases the material from epitaxial constraints, allowing its lattice and electronic structures to approach their intrinsic configurations, thereby underpinning the enhancement of key order parameters such as ferroelectricity, magnetism, and superconductivity. Concurrently, the interface decoupling process effectively mitigates defects and pinning effects caused by strong chemical bonds. This improvement in interface quality helps reduce charge scattering, thereby revealing the intrinsic charge transport behavior and quantum phenomena of materials in the low-dimensional limit with greater clarity.

Furthermore, the freestanding architecture enables novel mechanical responses in traditionally brittle materials. Through mechanisms such as domain switching and controlled buckling, efficient strain energy dissipation is achieved, ultimately leading to a paradigm shift from rigidity to flexibility and enabling potential applications in flexible electronics. In summary, delamination and transfer technologies have demonstrated significant potential as powerful tools for strain engineering and property control. By proactively designing the two core degrees of freedom — "interface" and "strain" — these techniques open expansive prospects for exploring novel material phases and functionalities in low-dimensional systems.

### 5. Conclusion and Outlook

Freestanding thin-film technology has emerged as a transformative platform that fundamentally reshapes materials design paradigms. As comprehensively reviewed in this work, diverse fabrication strategies—spanning physical delamination, chemical etching, and remote epitaxy— enable the liberation of high-quality films from substrate constraints, while advanced transfer techniques facilitate precise heterointegration across arbitrary material combinations. The resulting substrate-free architectures unlock suppressed intrinsic properties through strain relaxation and interface decoupling, while simultaneously introducing unprecedented mechanical flexibility. These capabilities have already enabled breakthrough applications in quantum computing, flexible electronics, and biomedical devices. However, the most profound impact of this technology lies not merely in property restoration, but in the emergence of entirely new design freedoms that transcend conventional epitaxial limitations. This concluding section examines these transformative opportunities alongside critical challenges and future directions.

## 5.1 Geometric Control: A Post-Growth Design Paradigm

Beyond restoring intrinsic properties through substrate liberation, freestanding thin-film technology uniquely enables a geometric design strategy impossible in conventional epitaxy: the independent control of stacking order, twist angles, and interlayer coupling. This capability transcends lattice-matched heteroepitaxy, where interfacial configurations are substrate-determined. The programmable manipulation of these geometric parameters—stacking registry, rotational misorientation, and interfacial strain—creates a new axis for engineering quantum materials. Unlike the continuous parameter space of chemical doping or applied fields, geometric control operates through discrete symmetry breaking and moiré potential modulation, generating emergent electronic structures fundamentally distinct from constituent layers. This approach establishes freestanding films not merely as substrate-free versions of epitaxial systems, but as an entirely new materials platform where structure-property relationships can be designed on demand through post-growth assembly rather than being predetermined during synthesis.

## 5.2 Remaining Challenges

Despite these advances, the pathway toward widespread implementation faces several critical bottlenecks requiring systematic solutions:

(1) Scalability and Process Control Limitations. While recent advances in sacrificial layers (e.g., $Sr_4Al_2O_7$) have improved dissolution kinetics, achieving uniform, defect-free release across wafer-scale areas remains challenging, requiring systematic optimization of etching rates, temperature gradients, and interfacial stress distribution for industrial-scale production.

(2) Transfer-Induced Contamination and Defects. Limited control over contamination, interfacial bubbles, and stress-induced deformation continues to compromise device performance. The formation of wrinkles, cracks, and residual substrate material during transfer remains a significant obstacle to achieving pristine interfaces required for high-performance quantum and electronic devices.

(3) Technological Generality and Material Compatibility. Current techniques remain material-specific, with strongly bonded systems and non-oxide materials requiring breakthroughs in interfacial energetics and fracture mechanics control.

**5.3 Future Directions and Transformative Potential**

Realizing the full transformative potential of this technology necessitates coordinated advances across multiple frontiers:

(1) Advanced Fabrication and Transfer Technologies. Future progress requires the development of: (i) next-generation sacrificial layers with enhanced epitaxial compatibility, reduced interfacial stress, and multi-functional capabilities (serving as both growth templates and rapid-release media); (ii) automated, damage-free transfer systems employing probe-based micro-positioning, thermally programmable interfaces, and vacuum-assisted integration to enable complex heterostructures with atomically sharp interfaces; and (iii) scalable manufacturing processes transitioning from laboratory demonstrations to industrial production.

(2) Strain-Gradient and Geometric Engineering. Exploring strain-gradient-induced emergent phenomena and programmable geometric control (stacking configurations, twist angles, dimensional confinement) represents a paradigm shift from passive property preservation to active functionality creation. The interplay between flexoelectric/flexomagnetic effects, topological band restructuring, and correlated quantum states in ultra-thin geometries offers unprecedented opportunities for designing reconfigurable electronic and spintronic devices.

(3) Multi-Physical Field Integration for Intelligent Systems. Integrating mechanical strain, electric/magnetic fields, and optical excitation will enable smart responsive systems with dynamically tunable properties, particularly promising for adaptive quantum sensors, neuromorphic computing platforms, and bio-inspired intelligent devices.

In conclusion, freestanding thin-film technology represents a paradigm shift in the design and manipulation of functional thin-film materials—evolving from a strategy of substrate liberation toward proactive design platforms for synergistic control of structural, mechanical, and quantum properties. As fabrication precision, transfer fidelity, and geometric engineering converge, this technology stands poised to redefine the boundaries of materials manipulation, opening pathways to devices and functionalities that bridge atomic-scale control with macroscopic performance. The journey from fundamental understanding of interfacial energetics to practical implementation embodies the quintessential challenge and opportunity of modern materials research.

**References**

[1] R. Peng, et al., Acta Phys. Sin. 72 (2023) 098502.

[2] A. Wang, et al., J. Univ. Sci. Technol. China 54 (2024) 0701.

[3] M. Konagai, et al., J. Cryst. Growth 45 (1978) 277-280.

[4] J. W. Matthews, J. Vac. Sci. Technol. 3 (1966) 133-145.

[5] A. Potts, et al., J. Phys.: Condens. Matter 2 (1990) 1807-1815.

[6] J. Zhang, et al., Nano Lett. 22 (2022) 7370-7377.

[7] G. Pérez-Mitta, et al., Sci. Adv. 10 (2024) eado4722.

[8] I. Shahine, et al., ACS Appl. Electron. Mater. 6 (2024) 2281-2288.

[9] J. Arokiaraj, et al., Appl. Phys. Lett. 88 (2006) 221901.

[10] J. Saleem, et al., Energy Rep. 9 (2023) 31-39.

[11] M. V. Kelso, et al., Science 364 (2019) 166-169.

[12] D. Li, et al., Energy Environ. Sci. 14 (2021) 424-436.

[13] H. Nie, et al., ACS Appl. Mater. Interfaces 8 (2016) 1937-1942.

[14] M. Zhou, et al., RSC Adv. 7 (2017) 49568-49575.

[15] J. Guo, et al., Nano Lett. 24 (2024) 1114-1121.

[16] S. Liang, et al., Appl. Phys. Lett. 95 (2009) 182509.

[17] W. W. Gao, et al., J. Appl. Phys. 117 (2015) 17C733.

[18] Q. Yang, et al., Acta Mater. 112 (2016) 216-223.

[19] N. B. Aetukuri, et al., Nat. Phys. 9 (2013) 661-666.

[20] H. J. Paik, et al., Appl. Phys. Lett. 107 (2015) 163101.

[21] N. F. Quackenbush, et al., Phys. Rev. B 94 (2016) 085105.

[22] T. D. Vethaak, et al., J. Appl. Phys. 129 (2021) 105104.

[23] Z. Feng, et al., Sci. Rep. 8 (2018) 4039.

[24] S. E. Kim, et al., Adv. Mater. Technol. 8 (2022) 2200878.

[25] X. Yang, et al., Nat. Commun. 15 (2024) 9281.

[26] Y. Li, et al., Adv. Mater. 34 (2022) 2203187.

[27] L. Dai, et al., Mater. Today Commun. 42 (2025) 111412.

[28] J. H. Kim, A. M. Grishin, Appl. Phys. Lett. 87 (2005) 033502.

[29] S. Kim, et al., Adv. Mater. 26 (2014) 7480-7487.

[30] K. Gu, et al., Nat. Nanotechnol. 17 (2022) 1065-1071.

[31] C. Jin, et al., Adv. Sci. 8 (2021) 2102178.

[32] S. Cai, et al., Nat. Commun. 13 (2022) 5116.

[33] J. Lin, et al., Adv. Mater. 37 (2025) 2414113.

[34] D. Ji, et al., Nature 570 (2019) 87-90.

[35] H. S. Lee, et al., Adv. Funct. Mater. 24 (2014) 6914-6921.

[36] C. K. Jeong, et al., Nano Res. 10 (2016) 437-455.

[37] B. B. Stogin, et al., Sci. Adv. 4 (2018) eaat3276.

[38] D. Li, et al., Nano Lett. 21 (2021) 4454-4460.

[39] R. Yang, et al., IEEE Trans. Nanotechnol. 18 (2019) 37-41.

[40] L. S. Farrar, et al., npj Quantum Mater. 5 (2020) 29.

[41] D. Li, et al., 24th Int. Workshop Oxide Electron. (iWOE 24), Chicago, IL, USA, 2017.

[42] L. J. Sandilands, et al., Phys. Rev. B 90 (2014) 081402.

[43] Y. Yu, et al., Nature 575 (2019) 156-163.

[44] S. Gao, et al., Phys. Rev. X 13 (2023) 041049.

[45] K. S. Novoselov, et al., Science 306 (2004) 666-669.

[46] L. H. Li, Y. Chen, Adv. Funct. Mater. 26 (2016) 2594-2608.

[47] J. Shim, et al., Science 362 (2018) 665-670.

[48] L. Zong, et al., Adv. Mater. 29 (2017) 1604691.

[49] N. E. Staley, et al., Phys. Rev. B 80 (2009) 184505.

[50] F. Dross, et al., Appl. Phys. A 89 (2007) 149-152.

[51] H. S. Kum, et al., Nature 578 (2020) 75-81.

[52] Y. Kim, et al., Nature 544 (2017) 340-343.

[53] L. Dai, et al., Nat. Commun. 13 (2022) 2990.

[54] S. R. Bakaul, et al., Nat. Commun. 7 (2016) 10547.

[55] S. Liu, et al., Nano-Micro Lett. 15 (2023) 131.

[56] F. M. Chiabrera, et al., Ann. Phys. 534 (2022) 2200084.

[57] X. Li, et al., Science 324 (2009) 1312-1314.

[58] H. Lu, et al., Nat. Commun. 5 (2014) 5518.

[59] M. T. Ghoneim, et al., Adv. Electron. Mater. 1 (2015) 1500045.

[60] C. Tan, et al., Chem. Rev. 117 (2017) 6225-6331.

[61] K. Kim, et al., Adv. Mater. 31 (2018) 1804690.

[62] F. Qing, et al., Nanoscale 12 (2020) 10890-10911.

[63] F. Liu, Prog. Surf. Sci. 96 (2021) 100626.

[64] W. Dong, et al., Adv. Mater. 36 (2023) 2303014.

[65] H. Liu, et al., ACS Nano 18 (2024) 11573-11597.

[66] Y. Huang, et al., Nat. Commun. 11 (2020) 2453.

[67] B. Zhang, et al., Nano-Micro Lett. 13 (2021) 39.

[68] S. Choo, et al., Sci. Adv. 10 (2024) eadq8561.

[69] Y. B. Park, et al., Adv. Mater. 18 (2006) 1533-1536.

[70] J. Kim, et al., Appl. Phys. A 122 (2016) 305.

[71] Y. Kim, et al., ACS Nano 16 (2022) 2399-2406.

[72] D. Wang, et al., Nanoscale 11 (2019) 9310-9318.

[73] S. Varshney, et al., ACS Nano 18 (2024) 6348-6358.

[74] D. Lu, et al., Nat. Mater. 15 (2016) 1255-1260.

[75] J. Zhang, et al., Science 383 (2024) 388-394.

[76] L. Han, et al., Adv. Mater. Interfaces 7 (2020) 1901604.

[77] S. S. Hong, et al., Science 368 (2020) 71-76.

[78] J. Wang, et al., Adv. Sci. 11 (2024) 2401657.

[79] F. An, et al., Adv. Funct. Mater. 30 (2020) 2003495.

[80] Y. Yue, et al., Adv. Mater. 36 (2024) 2313971.

[81] J. W. Matthews, A. E. Blakeslee, J. Cryst. Growth 27 (1974) 118-125.

[82] J. C. Lambropoulos, et al., J. Appl. Phys. 66 (1989) 4230-4242.

[83] W. Tian, et al., ECS J. Solid State Sci. Technol. 11 (2022) 046001.

[84] W. S. Wong, et al., Appl. Phys. Lett. 72 (1998) 599-601.

[85] M. Tomczyk, et al., J. Mater. Chem. C 5 (2017) 12529-12537.

[86] H. Yu, et al., Adv. Funct. Mater. 27 (2017) 1700461.

[87] J. H. Kwon, et al., ACS Appl. Mater. Interfaces 10 (2018) 15829-15840.

[88] Y. J. Ko, et al., ACS Appl. Mater. Interfaces 8 (2016) 6504-6511.

[89] H. Palneedi, et al., APL Mater. 5 (2017) 096111.

[90] S. S. Won, et al., Nano Energy 55 (2019) 182-192.

[91] K. Kerman, S. Ramanathan, J. Mater. Res. 29 (2014) 320-337.

[92] J. Bian, et al., Adv. Electron. Mater. 5 (2019) 1800900.

[93] F. Wang, et al., Adv. Mater. Technol. 8 (2022) 2201186.

[94] M. K. Kelly, et al., Appl. Phys. Lett. 69 (1996) 1749-1751.

[95] P. C. Wu, Y. H. Chu, J. Mater. Chem. C 6 (2018) 6102-6117.

[96] L. Tan, et al., 2007 8th Int. Conf. Electron. Packag. Technol., Shanghai, China, (2007) 1-5.

[97] W. Sun, et al., Adv. Funct. Mater. 32 (2022) 2111920.

[98] L. Tsakalakos, T. Sands, Appl. Phys. Lett. 76 (2000) 227-229.

[99] C. K. Jeong, et al., Energy Environ. Sci. 7 (2014) 4035-4043.

[100] K. I. Park, et al., Adv. Mater. 26 (2014) 2514-2520.

[101] I. French, et al., SID Symp. Dig. Tech. Pap. 36 (2005) 1634-1637.

[102] M. Byun, Thin Solid Films 663 (2018) 31-36.

[103] H. Y. Hsiao, et al., 2016 IEEE 18th Electron. Packag. Technol. Conf. (EPTC), Singapore, (2016) 43-46.

[104] K. Kim, et al., J. Mater. Chem. C 2 (2014) 2144-2149.

[105] S. J. Kim, et al., ACS Nano 10 (2016) 10851-10857.

[106] B. Lei, et al., Phys. Rev. Lett. 116 (2016) 077002.

[107] J. Shiogai, et al., Nat. Phys. 12 (2016) 42-46.

[108] X. Wang, et al., Physica C 474 (2012) 13-17.

[109] A. Koma, J. Cryst. Growth 201 (1999) 236-241.

[110] C.-H. Ma, et al., Appl. Phys. Lett. 108 (2016) 253104.

[111] Y.-H. Chu, npj Quantum Mater. 2 (2017) 67.

[112] J. Liu, et al., Adv. Electron. Mater. 4 (2018) 1700522.

[113] C. Zhang, et al., ACS Appl. Mater. Interfaces 13 (2021) 28442-28450.

[114] S. Liu, et al., ACS Appl. Mater. Interfaces 12 (2020) 54991-54999.

[115] S. W. Bedell, et al., J. Appl. Phys. 122 (2017) 025103.

[116] Y. Shin, et al., Nature Materials 23 (2024) 1411-1420.

[117] A. Sambri, et al., Adv. Funct. Mater. 30 (2020) 1909964.

[118] M. Bruel, Nucl. Instrum. Methods Phys. Res. B 108 (1996) 313-319.

[119] T. Izuhara, et al., Appl. Phys. Lett. 82 (2003) 616-618.

[120] R. T. Dahm, et al., ACS Appl. Mater. Interfaces 13 (2021) 12341-12346.

[121] R. Erlandsen, et al., Nano Lett. 22 (2022) 4758-4764.

[122] X. Zhang, et al., Nature 641 (2025) 98-105.

[123] S.-H. Bae, et al., Nat. Mater. 18 (2019) 550-560.

[124] H. Kum, et al., Nat. Electron. 2 (2019) 439-450.

[125] Y. Bitla, Y.-H. Chu, Nanoscale 12 (2020) 18523-18544.

[126] H. Kim, et al., Nat. Rev. Methods Primers 2 (2022) 40.

[127] Y. Qu, et al., ACS Appl. Mater. Interfaces 14 (2022) 2263-2274.

[128] W. Kong, et al., Nat. Mater. 17 (2018) 999-1004.

[129] H. Kim, et al., ACS Nano 15 (2021) 10587-10596.

[130] K. Qiao, et al., Nano Lett. 21 (2021) 4013-4020.

[131] K. Chung, et al., Science 330 (2010) 655-657.

[132] Y. Kobayashi, et al., Nature 484 (2012) 223-227.

[133] J. Kim, et al., Science 342 (2013) 833-836.

[134] S. A. Lee, et al., ACS Appl. Mater. Interfaces 9 (2017) 3246-3250.

[135] J. Jiang, et al., Nat. Commun. 10 (2019) 4145.

[136] J. Jeong, et al., Sci. Adv. 6 (2020) eaaz5180.

[137] J. Jeong, et al., ACS Appl. Nano Mater. 3 (2020) 8920-8930.

[138] J. Jeong, et al., Nano Energy 86 (2021) 106075.

[139] D. Du, et al., Nat. Commun. 12 (2021) 2494.

[140] Y. Kim, et al., Science 377 (2022) 859-864.

[141] Z. Chen, et al., Adv. Mater. 31 (2019) 1807345.

[142] P. Wang, et al., Appl. Phys. Lett. 116 (2020) 171905.

[143] J. Dong, et al., Nature 602 (2022) 606-611.

[144] Q. Gan, et al., Appl. Phys. Lett. 72 (1998) 978-980.

[145] H. W. Jang, et al., Phys. Rev. Lett. 101 (2008) 107602.

[146] Y. Zhang, et al., ACS Nano 11 (2017) 8002-8009.

[147] Y. Qi, et al., Nano Lett. 10 (2010) 524-528.

[148] Y. Qi, et al., Nano Lett. 11 (2011) 1331-1336.

[149] D. M. Paskiewicz, et al., Nano Lett. 16 (2015) 534-542.

[150] U. Cigane, et al., Micromachines 14 (2023) 1419.

[151] H. W. Jang, et al., Appl. Phys. Lett. 92 (2008) 062910.

[152] J. A. Rogers, MRS Bull. 39 (2014) 549-556.

[153] R. Takahashi, M. Lippmaa, ACS Appl. Mater. Interfaces 12 (2020) 25042-25049.

[154] S. Chen, et al., J. Appl. Phys. 133 (2023) 045303.

[155] S. R. Bakaul, et al., Adv. Mater. 29 (2017) 1605699.

[156] L. Shen, et al., Adv. Mater. 29 (2017) 1702411.

[157] Y. Shen, et al., Nat. Commun. 15 (2024) 4789.

[158] D. Pesquera, et al., Nat. Commun. 11 (2020) 3190.

[159] Y.-W. Chang, et al., Nanoscale Res. Lett. 15 (2020) 172.

[160] G. Dong, et al., Science 366 (2019) 475-479.

[161] S. G. Deutscher, E. Grunbaum, U.S. Patent 4,255,208, Mar. 10, 1981.

[162] A. K. Prodjosantoso, B. J. Kennedy, J. Solid State Chem. 168 (2002) 229-236.

[163] B. Lazic, et al., Solid State Sci. 11 (2009) 77-84.

[164] R. Vendamme, et al., Nat. Mater. 5 (2006) 494-501.

[165] G. Dong, et al., Adv. Mater. 32 (2020) 2004477.

[166] Y. Bourlier, et al., ACS Appl. Mater. Interfaces 12 (2020) 8466-8474.

[167] S. Moon, et al., Sci. Rep. 6 (2016) 30107.

[168] G. Cha, et al., Chem. Asian J. 11 (2016) 789-797.

[169] H. Elangovan, et al., ACS Nano 14 (2020) 5053-5060.

[170] H. Zhong, et al., Adv. Mater. 34 (2022) 2109889.

[171] R. Xu, et al., Adv. Mater. 35 (2023) 2210562.

[172] H. Peng, et al., Adv. Funct. Mater. 32 (2022) 2111907.

[173] C. Lu, et al., ACS Nano 18 (2024) 5374-5382.

[174] J. Li, et al., ACS Appl. Nano Mater. 7 (2024) 9696-9702.

[175] P. Salles, et al., Adv. Funct. Mater. 33 (2023) 2304059.

[176] K. Han, et al., ACS Appl. Mater. Interfaces 13 (2021) 16688-16693.

[177] P. Salles, et al., Adv. Mater. Interfaces 8 (2021) 2001643.

[178] G. Dong, et al., Adv. Mater. 34 (2022) 2108419.

[179] Y. Li, et al., Adv. Mater. 34 (2022) 2106826.

[180] H. Liu, et al., ACS Catal. 12 (2022) 4119-4124.

[181] Q. Wang, et al., Appl. Catal. B 317 (2022) 121781.

[182] L. Han, et al., Nano Lett. 23 (2023) 2808-2815.

[183] R. Qiu, et al., Thin Solid Films 773 (2023) 139820.

[184] L. Nian, et al., Adv. Mater. 36 (2024) 2307682.

[185] L. Xiang, et al., Phys. Rev. Mater. 8 (2024) 054417.

[186] K. Gu, et al., Adv. Funct. Mater. 30 (2020) 2001236.

[187] Z. Lu, et al., APL Mater. 8 (2020) 051105.

[188] B. Peng, et al., Sci. Adv. 6 (2020) eaba5847.

[189] Q. Wang, et al., Crystals 10 (2020) 733.

[190] K. Gu, et al., Adv. Mater. (2025) 2505707.

[191] K. Han, et al., Sci. Adv. 5 (2019) eaaw7286.

[192] N. Liu, et al., Clin. Transl. Med. 14 (2024) e1641.

[193] S. Yan, et al., Adv. Mater. 36 (2024) 2402916.

[194] Y. Hong et al., Nat. Commun. 15 (2024) 5030.

[195] S. Liu et al., Nat. Commun. 15 (2024) 10136.

[196] P. Singh, et al., ACS Appl. Electron. Mater. 1 (2019) 1269-1274.

[197] Y. Lee, et al., Nat. Synth. 4 (2025) 573-581.

[198] S. Liu, et al., Adv. Sci. 9 (2022) 2106030.

[199] Z. Zhang, et al., Nat. Commun. 14 (2023) 4094.

[200] Y. Cao, et al., Nature 556 (2018) 80-84.

[201] Y. Cao, et al., Nature 556 (2018) 43-50.

[202] J. M. Park, et al., Nature 590 (2021) 249-255.

[203] R. Ribeiro-Palau, et al., Science 361 (2018) 690-693.

[204] J. Yuan, et al., Chin. Phys. B 31 (2022) 08730.2

[205] M. Onodera and T. Machida, Sens. Mater. 35 (2023) 1929-1939.

[206] C.R. Woods, et al., Nat. Commun. 7 (2016) 10800.

[207] D. Wang, et al., Phys. Rev. Lett. 116 (2016) 126101.

[208] Y. Cheng, et al., Chin. Phys. B 28 (2019) 107304.

[209] N. Pryds, et al., APL Mater. 12 (2024) 010901.

[210] S. Esmaeili, et al., Opt. Commun. 285 (2012) 2243-2246.

[211] H. Meng, et al., Phys. Rev. B 107 (2023) 035109.

[212] D. Rong, et al., Nano Lett. 25 (2025) 5965-5973.

[213] J. Shen, et al., ACS Appl. Mater. Interfaces 14 (2022) 50386-50392.

[214] Z. Zhou, et al., Nature 621 (2023) 499-505.

[215] M. Tanaka, et al., Nature 638 (2025) 99-105.

[216] S. Lee, et al., Sci. Adv. 10 (2024) eadq0293.

[217] P.-C. Wu, et al., Nat. Commun. 13 (2022) 2565.

[218] Y. Wang, et al., Nano Res. 16 (2023) 7829-7836.

[219] L. Han, et al., Nature 603 (2022) 63-67.

[220] H. Sun, et al., Nat. Commun. 13 (2022) 4332.

[221] Z. Zhang, et al., Adv. Mater. 34 (2022) 2200864.

[222] M. Mojtabavi, et al., Small 18 (2022) 2105857.

[223] C. E. Ren, et al., J. Phys. Chem. Lett. 6 (2015) 4026-4031.

[224] L. Ding, et al., Nat. Commun. 9 (2018) 155.

[225] H. E. Lee, et al., Adv. Mater. 30 (2018) 1800649.

[226] H. Kim, et al., J. Appl. Phys. 130 (2021) 174901.

[227] Q. Paduano, et al., J. Mater. Res. 31 (2016) 2204-2213.

[228] L. Liu, et al. Nano Lett. 25 (2025) 14213-14221.

[229] Q. Wang, et al., 2D Mater. 2 (2015) 044012.

**Table 1 Freestanding Thin-Film Fabrication Methods**

| Category | Method | Principle | Typical Materials/ Structures | Key Advantages | Major Limitations |
|---|---|---|---|---|---|
| Physical Delamination | **Laser Lift-off** | Selective energy deposition at interface → Vaporization → Shock wave separation | GaN/sapphire PLZT/MgO | - Non-contact processing<br>- Recyclable transparent substrates<br>- High selectivity | - Requires transparent substrate and absorbing layer<br>- Ultrathin films prone to cracking<br>- Expensive equipment |
| | **Mechanical Exfoliation/ Spalling** | Weak interface/stress layer → External force triggers cracking | FeSe/PDMS $La_{1-x}Sr_xMnO_3$/mica BCZT/mica | - Chemical-free<br>- Substrate reusable<br>- Ideal for layered materials (like mica or GaN) | - High interface sensitivity<br>- Substrate residue<br>- Requires engineered weak interface |
| | **Remote Epitaxy vdW Lift-off** | 2D-coated substrate epitaxy → vdW interface enables lift-off | GaN/graphene $SrTiO_3/SiO_2$/Si PMN-PT/$CoFe_2O_4$ | - Overcomes lattice matching<br>- Substrate recyclable<br>- Atomically flat interface | - Requires ionic substrate (>30%)<br>- 2D layer quality control<br>- Dry transfer required |
| | **Other Physical Lift-off** | Ion beam sputtering etching, ultrasonic cavitation overcomes interlayer forces | $La_{1-x}Sr_xMnO_3/BaTiO_3/CeO_2$/YSZ heterostructure | - Non-thermal process (suitable for thermally sensitive materials)<br>- Atomic-level precision (ion beam) | - Low efficiency<br>- Limited material scope |
| Chemical Lift-off | **Substrate Etching** | Selective dissolution of substrate | $SrRuO_3/SrTiO_3$ (HF) $BiFeO_3$/Si (HF) PZT/MgO ($H_3PO_4$) | - Simple equipment<br>- High dissolution selectivity (>100:1)<br>- Releases epitaxial strain | - Substrate not reusable<br>- Film fragility<br>- Solvent specificity |
| | **Sacrificial Layer Etching** | Selective dissolution of sacrificial interlayer | - **Water-soluble:** $Sr_3Al_2O_6(BiFeO_3)$ $Sr_4Al_2O_7$(nickelates)<br>- **Acid-soluble**: $La_{1-x}Sr_xMnO_3$, $YBa_2Cu_3O_{7-x}$<br>- **Special etchant**: $SrRuO_3(NaIO_4)$ | - Compatible with complex oxides<br>- Strain release enhances properties<br>- Pre-strain design possible | - Limited material scope (water-soluble)<br>- Acid residue requires annealing<br>- Interface residue risk |

**Table 2 Classification of Sacrificial Layer Etching**

| Type | Sacrificial Material | Etchant | Key Impacts | Examples |
|---|---|---|---|---|
| **Water-soluble** | Alkaline earth aluminates ($Ba_3Al_2O_6$, $Sr_{3-x}Ca_xAl_2O_6$, $Sr_4Al_2O_7$) | Deionized water | -Slow $Sr_3Al_2O_6$ dissolution and cracks<br>-$Sr_4Al_2O_7$: 10× faster dissolution | Freestanding thin films: $BiFeO_3$[33,188], $BaTiO_3$ [78,160,165,186,189], LPCMO [185], $La_{0.8}Sr_{0.2}NiO_2$ [193], $SrTiO_3$ [166] |
| | Alkaline earth oxides (SrO, BaO) | Deionized water | -Rapid dissolution,<br>-MBE-compatible | -SrO: $SrTiO_3$ [73]<br>-BaO: $BaTiO_3$ [153], $Al_2O_3$ [154] |
| | $SrVO_3$ | Deionized water | Eco-friendly process | $SrTiO_3$ [166] |
| **Acid-soluble** | $La_{1-x}Sr_xMnO_3$ | KI+HCl | Acid residue requires annealing | PZT [54,155], $BaTiO_3$ [169], $LiFe_5O_8$ [156], $Hf_{0.5}Zr_{0.5}O_2$[157,170], $NaNbO_3$ [171] |
| | $YBa_2Cu_3O_{7-x}$ | HCl | Acid residue requires annealing | $La_{1-x}Sr_xMnO_3$ [159] |
| | $SrCoO_{2.5}$ | Acetic acid (vinegar) | Eco-friendly process | $SrRuO_3$ [172] |
| | $Ga_{0.3}Al_{0.7}As$, AlAs | HF | -Simple & low-cost<br>-Limited material scope | GaAs [3,167] |
| **Special solvent** | $SrRuO_3$ | Oxidant ($NaIO_4$) | Avoids acid damage | $La_{1-x}Sr_xMnO_3$ [158] |
| | $TiO_2$ | $H_2O_2$ solution | Eco-friendly process | $TiO_2$ nanotube arrays [168] |
| | MgO | $(NH_4)_2SO_4$ solution | Mild etching conditions | $CoFe_2O_4$ [146] |

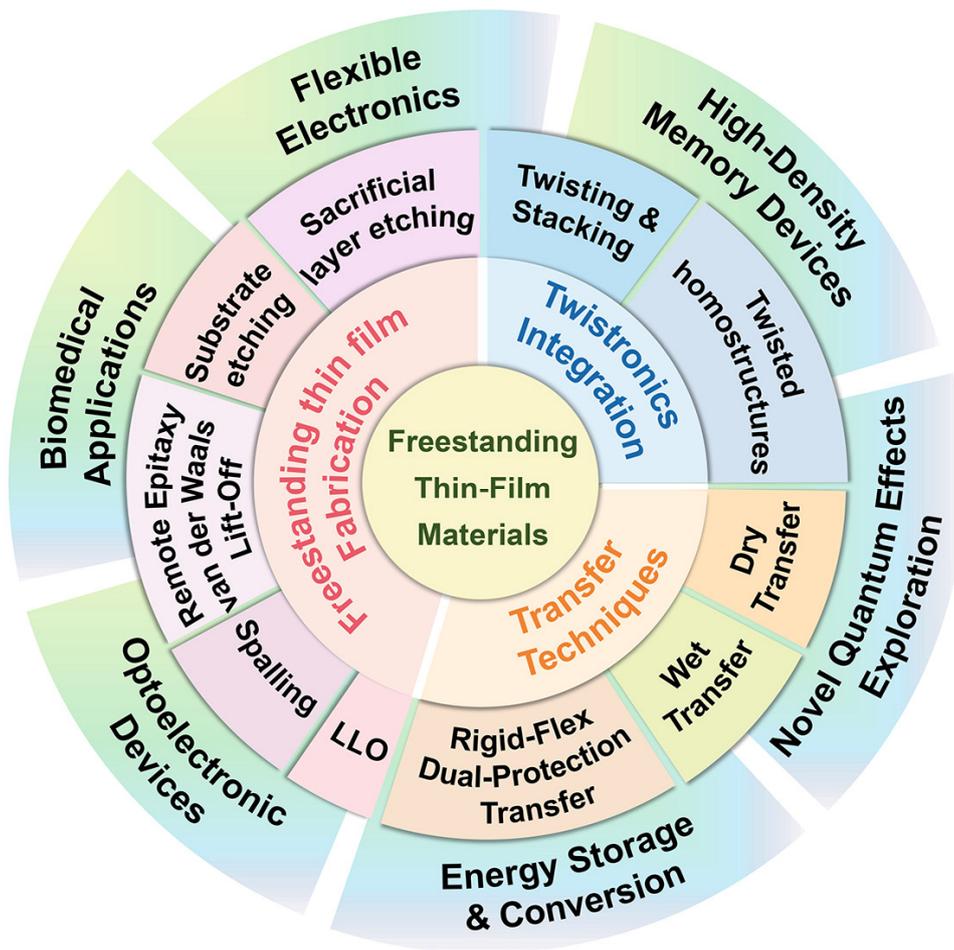

**Fig. 1.** Schematics of the fabrication, transfer, integration techniques and applications of freestanding thin-film materials.

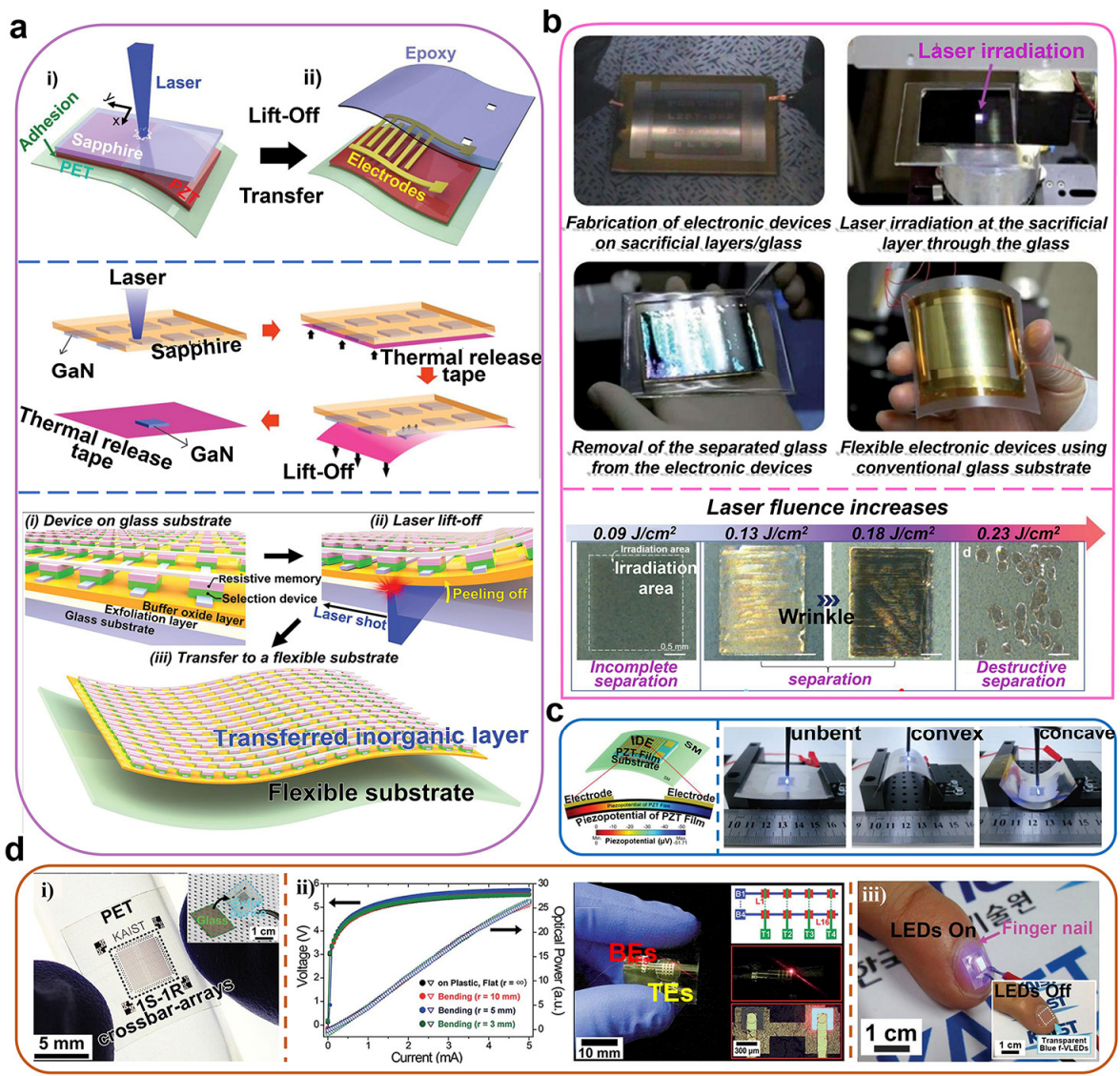

**Fig. 2. Laser lift-off (LLO) techniques.**

**(a)** From top to bottom: Fabrication process of high-efficiency flexible large-area PZT thin film nanogenerators via LLO [100]; selective laser lift-off mechanism [70]; fabrication of flexible crossbar memory on plastic substrates using inorganic-based laser lift-off (ILLO) [29]. **(b)** top: Generic process flow for LLO of organic light-emitting diodes (LEDs) [104]; bottom: optical micrographs of GaN films delaminated from sapphire at varied laser fluences [97]. **(c)** Simulated piezopotential distribution in freestanding PZT films between adjacent electrodes (left, model dimensions match experimental setup) [35]; violet-blue electroluminescence of separated GaN-LED devices under bending deformation (right) [97]. **(d)** Applications of freestanding films obtained by LLO: i) Flexible RRAM device on plastic substrate. Inset: Detachment process from glass after laser irradiation [29]; ii) Left panel: Luminance–current–voltage (L–I–V) characteristics at flat state and bending radii of 10/5/3 mm. Right panel: Crossbar-structured flexible vertical LED (f-VLED) array with activated unit. Top inset: Array architecture; middle inset: Activated f-VLED in dark room; bottom inset: Optical micrograph of illuminated f-VLED [99]; iii) GaN blue f-VLED array attached to human fingernail. Inset: Transparent array in off-state [225].

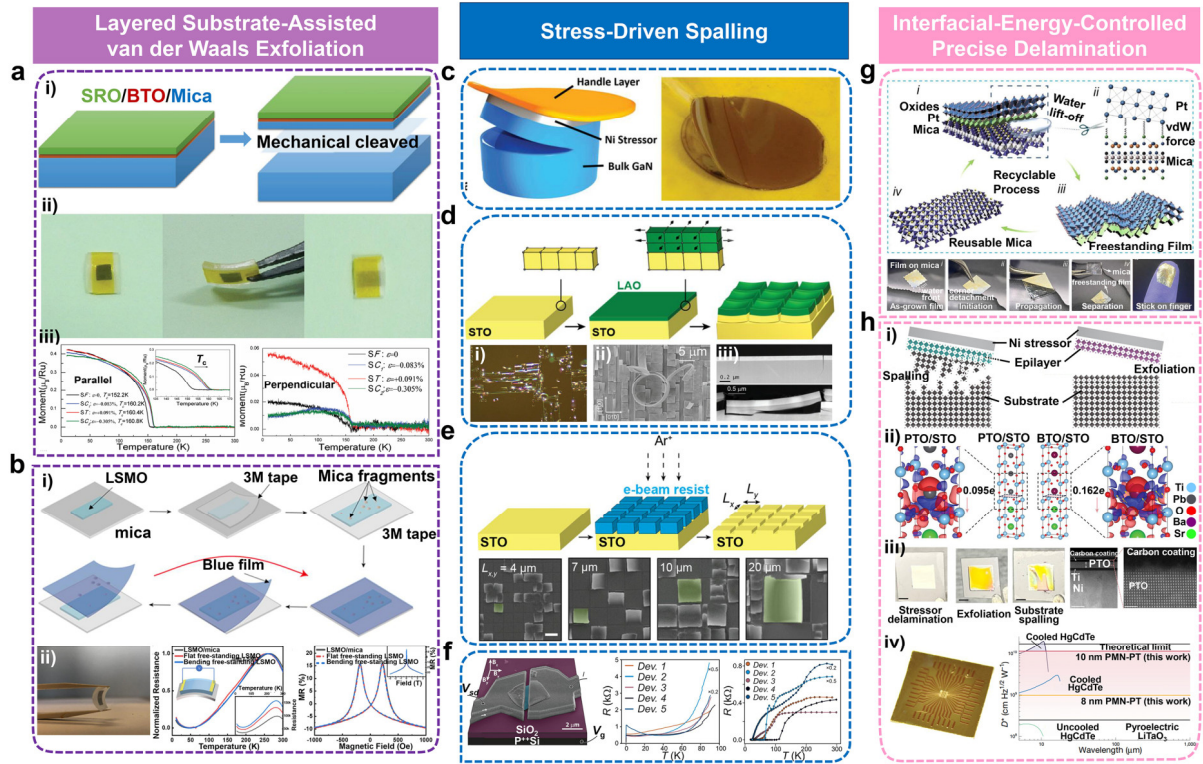

**Fig. 3. Mechanical exfoliation/Spalling techniques.**

**(a)** Mechanical exfoliation process, phenomena, and magnetic characterization of the freestanding SrRuO$_3$/BaTiO$_3$ heterostructure [112]. **(b)** i) Schematic of improved mechanical exfoliation for La$_{1-x}$Sr$_x$MnO$_3$ (adding blue-tape steps to remove residual mica vs. (a)). ii) Flexibility of freestanding La$_{1-x}$Sr$_x$MnO$_3$ sample (left); Normalized resistance-temperature (*R-T*) curves for three samples (middle); Magnetoresistance curves (right) [113]. **(c)** Spalling driven by external stress: Controlled Spalling of bulk GaN substrates left); Image of 2-inch GaN wafer during spalling (right) [115]. **(d)** Spalling driven by internal stress and spontaneous spalling behavior. Top panel: Schematic of spontaneous membrane spalling for LaAlO$_3$/SrTiO$_3$. Bottom panels: i) Optical image of an 80 nm-thick sample surface (brown area: uniform LaAlO$_3$ film; iridescent flakes: delaminated multilayer regions showing optical interference). ii) Scanning electron microscopy (SEM) image of a typical thick sample surface, showing membranes with lateral dimensions of 2–3 μm and rare helical tapes up to tens of micrometers. iii) Low-resolution cross-sectional electron microscopy images: Upper – 80 nm-thick LaAlO$_3$ sample in phase #2 (vertical crack development); Lower – 180 nm thick LaAlO$_3$ film in phase #3 (surface fracturing into bilayer membranes) [117,120]. **(e)** Internal stress lift-off with size control via surface patterning. Left: Schematic of surface patterning by Ar$^+$-ion milling; Right: Representative SEM images demonstrating precise size control in LaAlO$_3$ spalling, showing patterned dimensions ($L_{x,y}$ = 4–20 μm). Scale bar: 5 μm [120]. **(f)** Performance of freestanding LaAlO$_3$/SrTiO$_3$ membrane device. Left: Artificially colored SEM image of freestanding LaAlO$_3$/SrTiO$_3$ device on p$^{++}$-type Si/SiO$_2$ substrate (as backgate); Middle: Two-terminal resistance vs. temperature showing metallic behavior; Right: Superconducting transitions in low-*T* region of middle plot [121]. **(g)** vdW stripping of piezoceramic thin films. Top: Schematics of the vdW release process: i) Waterfront penetration at the mica/piezoceramic-Pt interface. ii) Microscopic view: Pt layer and mica sheets bonded via weak vdW forces, severable by external force (e.g., water). iii) Freestanding thin film obtained after complete interfacial water penetration. iv) Mica recyclable due to the non-destructive process. Bottom: Photographs of the thin film fabrication process (Left) and the freestanding Ba$_{0.85}$Ca$_{0.15}$Zr$_{0.1}$Ti$_{0.9}$O$_3$ film (with supporting EVA layer and bottom Pt electrode) on a gloved finger [55]. **(h)** Atomic Lift-off (ALO) Technology and Applications i) Schematics of two peeling modes: Left: Spalling - Non-uniform peeling with substrate damage; Right: Exfoliation - Atomic-level precision interface separation. ii)

Charge transfer visualization at PbTiO$_3$/SrTiO$_3$ (left) and BaTiO$_3$/SrTiO$_3$ (right) interfaces, elucidating Pb-induced chemical weakening mechanism in ALO. iii) Optical Microscopy images of peeling modes for PbTiO$_3$/STO with Ni stressor thicknesses (Ni delamination/PbTiO$_3$ delamination/substrate spalling); TEM images of freestanding PbTiO$_3$ (uniform thickness/atomically smooth surface). iv) ALO application in cooling-free far-infrared (FIR) imaging: Schematic of FIR imaging arrays (Left); Specific detectivity comparison of thermally isolated pyroelectric detectors, demonstrating performance superiority of intrinsic devices across the full FIR spectrum (Right) [122].

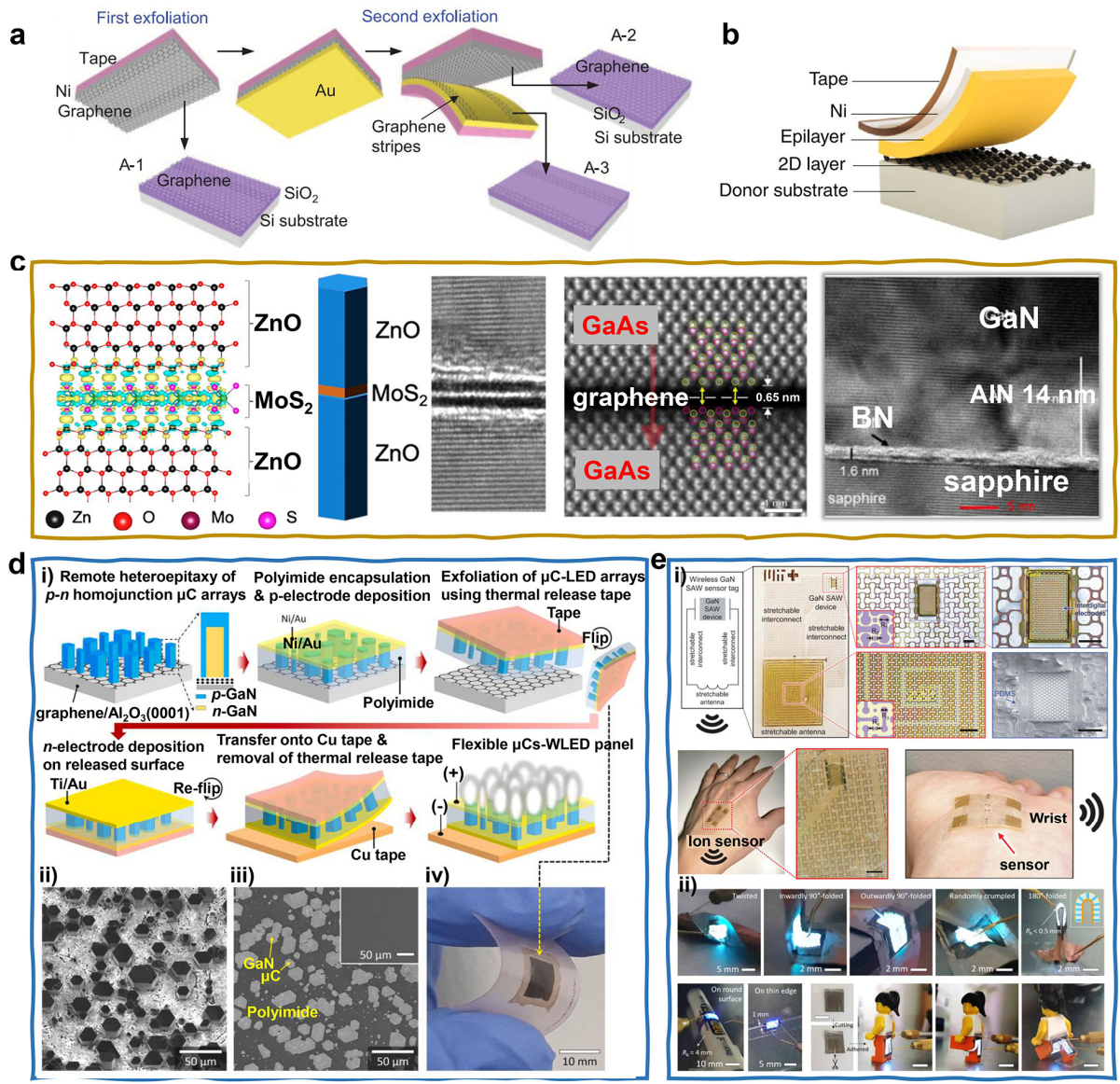

**Fig. 4. Remote epitaxy van der Waals (vdW) lift-off techniques.**
**(a)** Schematic of a method for removing double-layer stripes from graphene exfoliated from a SiC substrate [133]. **(b)** Schematics of epitaxial lift-off techniques using 2D material-assisted layer [124]. **(c)** First-principles calculations and multi-system transmission electron microscopy (TEM) characterizations of remote epitaxy. Left to right: Side view of First-principles calculation of ZnO/MoS$_2$/ZnO atomic configuration and charge density distribution of ZnO/monolayer MoS$_2$/ZnO heterostructure [71]; Cross-sectional TEM images of ZnO nanorod (NR)/ MoS$_2$/ZnO crystal, obtained by focusing on MoS$_2$ [71]; Cross-sectional high-angle annular dark-field scanning TEM (HAADF-STEM) image near the interface. The magenta and green circles indicate the Ga and the As atoms, respectively. The yellow arrows indicate the remote alignment of Ga and As atoms through graphene [226]; Cross-sectional high-resolution TEM (HR-TEM) image showing GaN/AlN insertion layer and AlGaN barrier [227]. **(d)** Remote epitaxy of GaN *p–n* homojunction microcolumn (μC) arrays on graphene-coated *c*-Al$_2$O$_3$ for flexible light-emitting diodes (LED) fabrication. i) Schematic illustration depicting fabrication procedures for flexible μCs-LED via remote epitaxy. ii) Tilt-view scanning electron microscopy (SEM) image of GaN *p–n* junction μCs-LED arrays on graphene/*c*-Al$_2$O$_3$ substrate. iii) Plan-view SEM image of released bottom surface of μCs-LED arrays encapsulated with polyimide filler. Inset SEM image displays surface of wafer after the exfoliation process. iv) Photograph of polyimide-encapsulated μCs exfoliated from substrate using thermal release tape [138]. **(e)** i) Schematic illustration and optical images of the wireless GaN surface acoustic wave (SAW) electronic skins (e-skin) strain sensor (top). Photographs and microscopy images of an e-skin attached on the back of a hand and wrist, respectively (bottom) [140]. ii) Diversely deformable microrod (MR) LEDs. Top: A series of photographs of cyan MR LED ($\lambda$ = 500 nm) deformed in various shapes. Left two (bottom): Photographs of blue MR LED ($\lambda$ = 450 nm) mounted on various surfaces. Right five (bottom): Photographs of 10 mm by 10 mm MR LED ($\lambda$ = 450 nm) tailored to be fitted to two back legs of a minifigure. (left two panels); photographs of LED-adhered LEGO minifigure with different leg postures (right three panels). Scale bars: 10 mm [136].

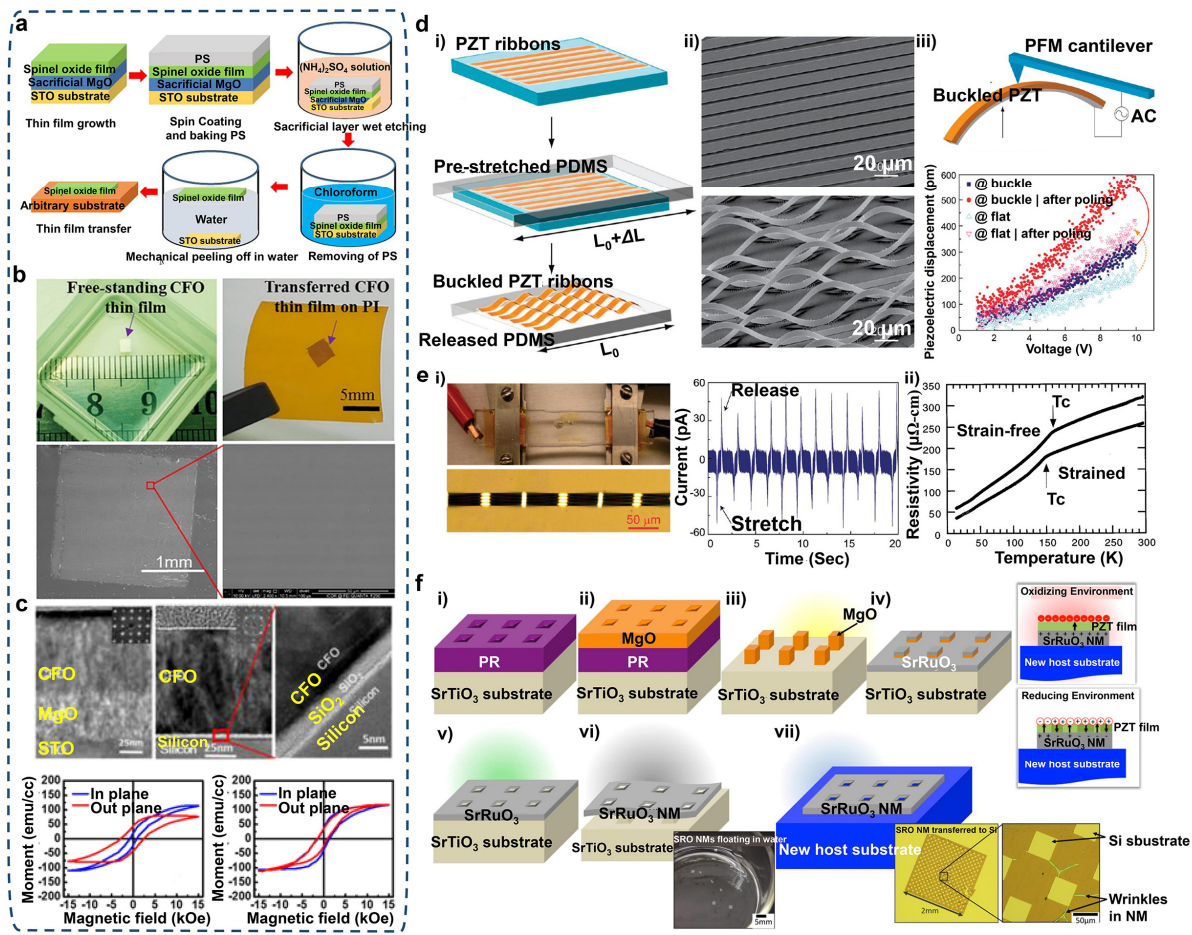

**Fig. 5. Substrate etching techniques.**
(a) Schematic illustration of the fabrication and transfer processes of freestanding $CoFe_2O_4$ thin films onto flexible substrates [146]. (b) Photographs and SEM images of freestanding $CoFe_2O_4$ thin films [146]. (c) Cross-sectional TEM images of as-grown $CoFe_2O_4$/MgO/$SrTiO_3$ multilayer films (left) and transferred freestanding $CoFe_2O_4$ thin films on silicon (right); paired with their corresponding in-plane and out-of-plane M–H hysteresis loops measured at room temperature [146]. (d) Formation of wavy/buckled PZT ribbons. i) Fabrication and transfer of patterned freestanding PZT onto prestrained PDMS, forming wavy/buckled structures upon strain relaxation. ii) SEM images of PZT ribbons printed onto PDMS: unbuckled at zero prestrain (top), buckled with prestrain (bottom). iii) Piezoelectric response via local probing in buckled ribbons [148]. (e) i) Enhanced energy conversion efficiency of stretched wavy/buckled freestanding PZT ribbons [148] and ii) improved electrical performance of strain-relaxed freestanding $SrRuO_3$ thin films. Arrows indicate Curie temperature $T_C$ [144]. (f) Fabrication of $SrRuO_3$ nanomembranes (NMs) via substrate pre-patterning, selective $SrRuO_3$ epitaxial growth, and selective etching. Insets: freestanding $SrRuO_3$ optical micrographs and schematics of electrochemical response for PZT heterostructures incorporating freestanding $SrRuO_3$ NMs under varied conditions [149].

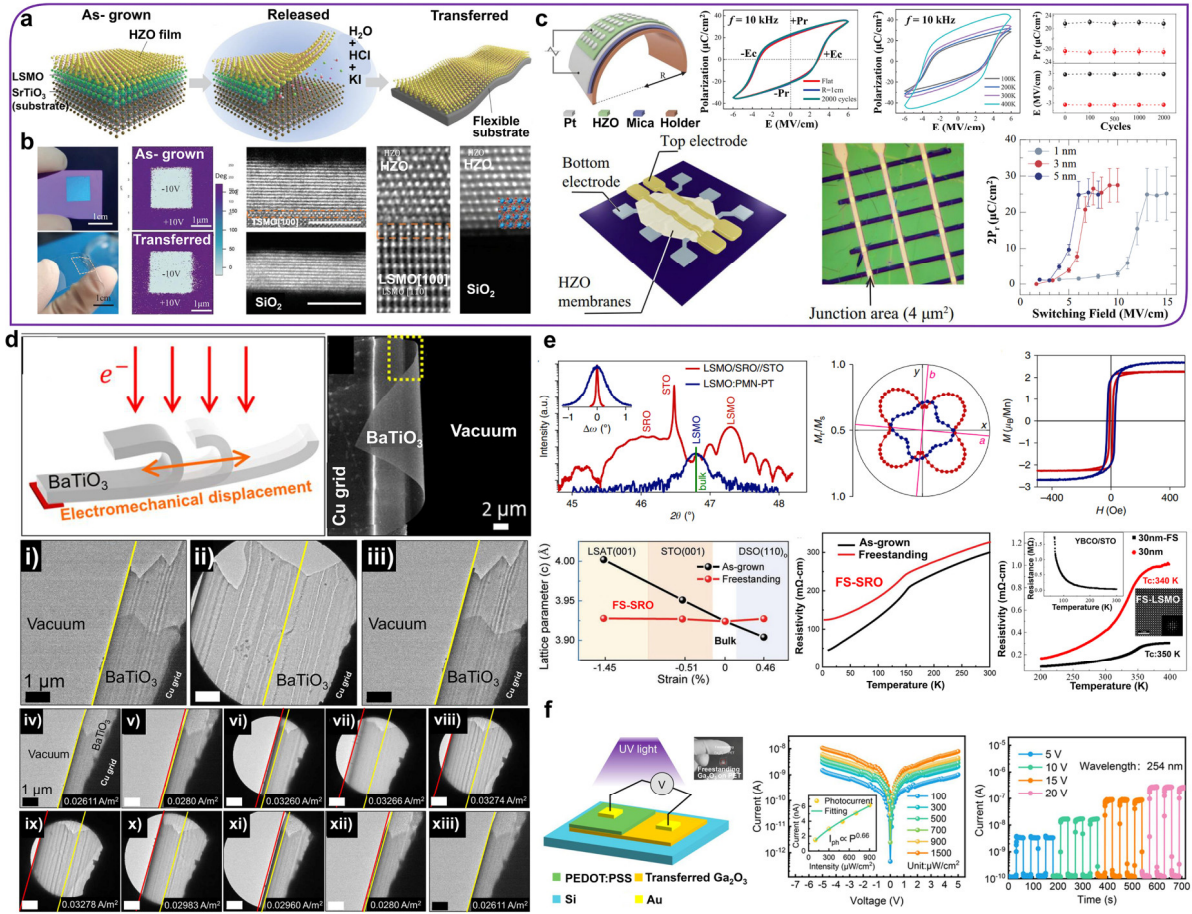

**Fig. 6. Water-insoluble sacrificial layer etching techniques.**
**(a)** Preparation and transfer of freestanding $Hf_{0.5}Zr_{0.5}O_2$ films using a $La_{0.8}Sr_{0.2}MnO_3$ sacrificial layer [170]. **(b)** Photograph of a 10 nm-thick freestanding $Hf_{0.5}Zr_{0.5}O_2$ film transferred to different substrates; along with out-of-plane piezoresponse force microscopy (PFM) phase images (Scale bar: 1 µm) and cross-sectional high-angle annular dark-field images (Scale bar: 5 nm) of the as-grown $Hf_{0.5}Zr_{0.5}O_2$ film on $La_{0.8}Sr_{0.2}MnO_3/SrTiO_3$ and the freestanding $Hf_{0.5}Zr_{0.5}O_2$ film transferred to $Pt/SiO_2$ [170]. **(c)** System test of the ferroelectric properties of the freestanding $Hf_{0.5}Zr_{0.5}O_2$ capacitor, fully verifying the robustness, stability and application potential of the freestanding film [157,170]. **(d)** Schematic and SEM micrographs of the folded freestanding $BaTiO_3$ thin film in its native state, along with its electromechanical rolling and unfolding response under an electron-beam-induced field [169]. **(e)** Lattice parameters, magnetic properties, and electrical properties of various freestanding FS thin films recover to the level comparable to bulk materials [158,159,172]. **(f)** Photocurrent response of the photodetector device based on the flexible freestanding *β*-$Ga_2O_3$ thin film under various illumination intensities and bias voltages [173].

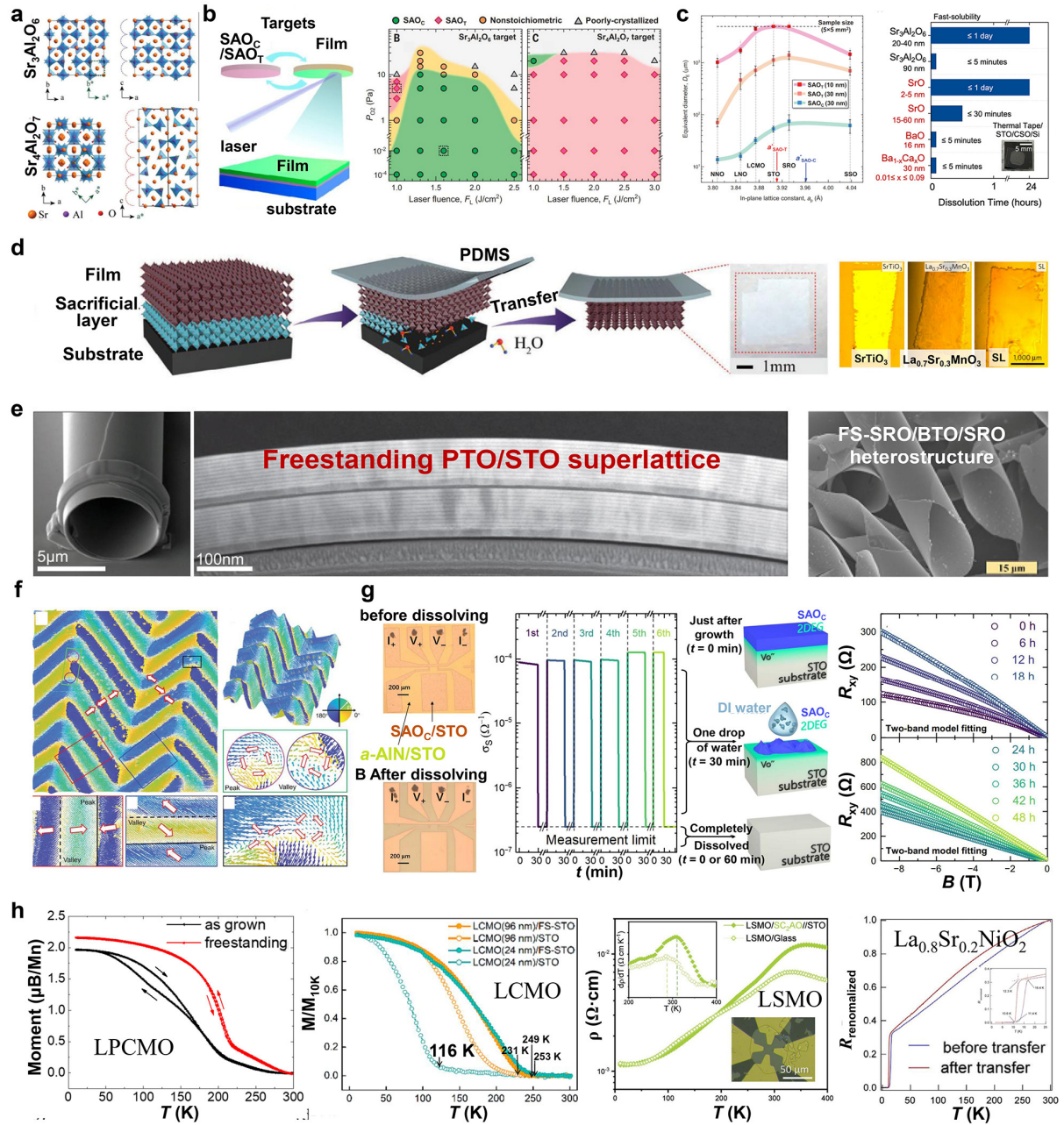

**Fig. 7. Water-soluble sacrificial layer etching techniques.**
**(a)** Schematics of the crystal structures of $Sr_3Al_2O_6$ and $Sr_4Al_2O_7$ [184]. **(b)** Schematic of pulsed laser deposition (PLD) growth for water-soluble sacrificial layers $Sr_3Al_2O_6$/ $Sr_4Al_2O_7$ and their growth conditions [74,184]. **(c)** Factors affecting freestanding film dimensions and sacrificial layer etching time: in-plane lattice parameters and sacrificial layer thickness, respectively [73,75]. **(d)** Schematic of the exfoliation and transfer process of freestanding thin films via sacrificial layer etching and with examples of some freestanding thin films [73,74]. **(e)** SEM and annular dark-field scanning transmission electron microscopy images of superelastic freestanding superlattice ($PbTiO_3$/$SrTiO_3$ tube) and heterostructure films ($SrRuO_3$/$BaTiO_3$/$SrRuO_3$), showing spontaneous curling [179,181]. **(f)** MATLAB-based conversion of vertical/lateral piezoresponse force microscopy (V-PFM/L-PFM) images of wrinkled freestanding $BaTiO_3$ films into vector contours, confirming periodic polarization topology induced by ultralow strain (≈0.5%). Vortex pairs (VPs) and vortex-antivortex pairs (VAPs) localize at wrinkle inflection points, where the strain gradient is maximal. Long-range ordered H–H (head-to-head) and T–T (tail-to-tail) ferroelectric domains are observed at peaks and valleys, respectively [78]. **(g)** Time-dependent sheet conductance ($\sigma_s$) of water-dissolvable recyclable devices based on a 20-nm $Sr_3Al_2O_6$/$SrTiO_3$ (001) heterointerface (300 K, ambient), demonstrating metal-insulator transition (MIT) induced by deionized water exposure time (t). Hall resistance evolves from nonlinear to linear with increasing $t$ [191]. **(h)** Enhanced/maintained magnetic and electrical properties of FS films after substrate-strain release. Curie temperature ($T_C$) of freestanding $(La_{2/3}Pr_{1/3})_{5/8}Ca_{3/8}MnO_3$ (LPCMO), freestanding $La_{0.7}Ca_{0.3}MnO_3$, and freestanding $La_{0.7}Sr_{0.3}MnO_3$ thin films increases compared to their as-grown states, approaching bulk values; superconductivity is preserved in freestanding $La_{0.8}Sr_{0.2}NiO_2$ thin film [15,175,185,193].

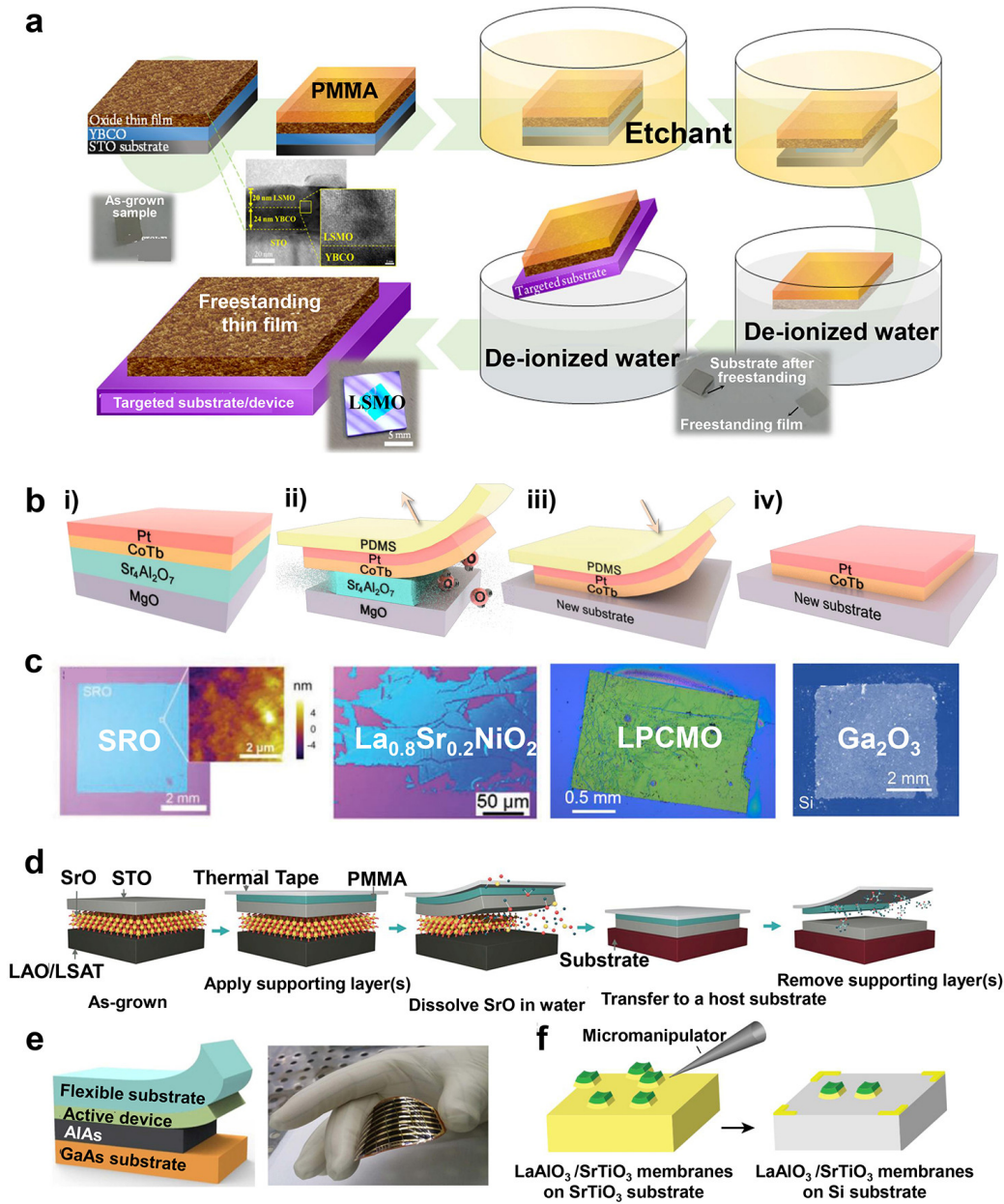

**Fig. 8. Transfer techniques for freestanding thin films.**

**(a)** Schematic and optical photographs of freestanding and wet transfer processes for the $La_xSr_{1-x}MnO_3$ thin film [159]. **(b)** Schematic illustration of freestanding thin film fabrication and dry transfer processes of the CoTb thin film [228]. **(c)** Optical photographs of dry- and wet-transferred thin films on silicon substrates [172,173,185,193]. (d) Schematic illustration of the rigid-flex dual-protection transfer process for the $SrTiO_3$ thin film [73]. **(e)** Schematic diagram of the simple and fast metal stress-assisted transfer lift-off [167]. **(f)** Schematic illustration of the controllable manipulation and transfer of $LaAlO_3/SrTiO_3$ thin films onto a silicon substrate using a micromanipulator needle [120].

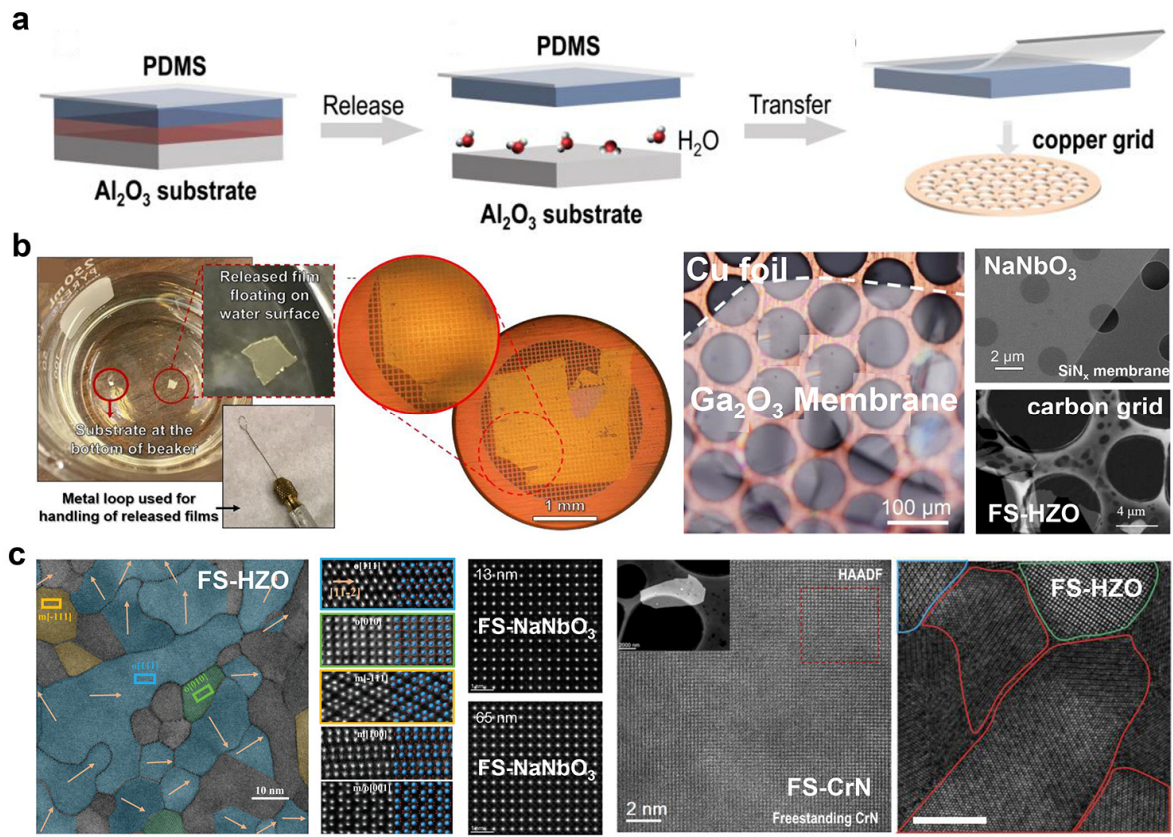

**Fig. 9. Transfer-enabled planar-view TEM technique.**
**(a)** Schematic of the fabrication process of freestanding thin films for STEM imaging [154]. **(b)** Optical microscopy images of freestanding thin films transferred onto TEM holey grids [169-171,173]. **(c)** Planar-view HAADF-STEM images of representative freestanding thin films fabricated via the developed technique [157,170,171].

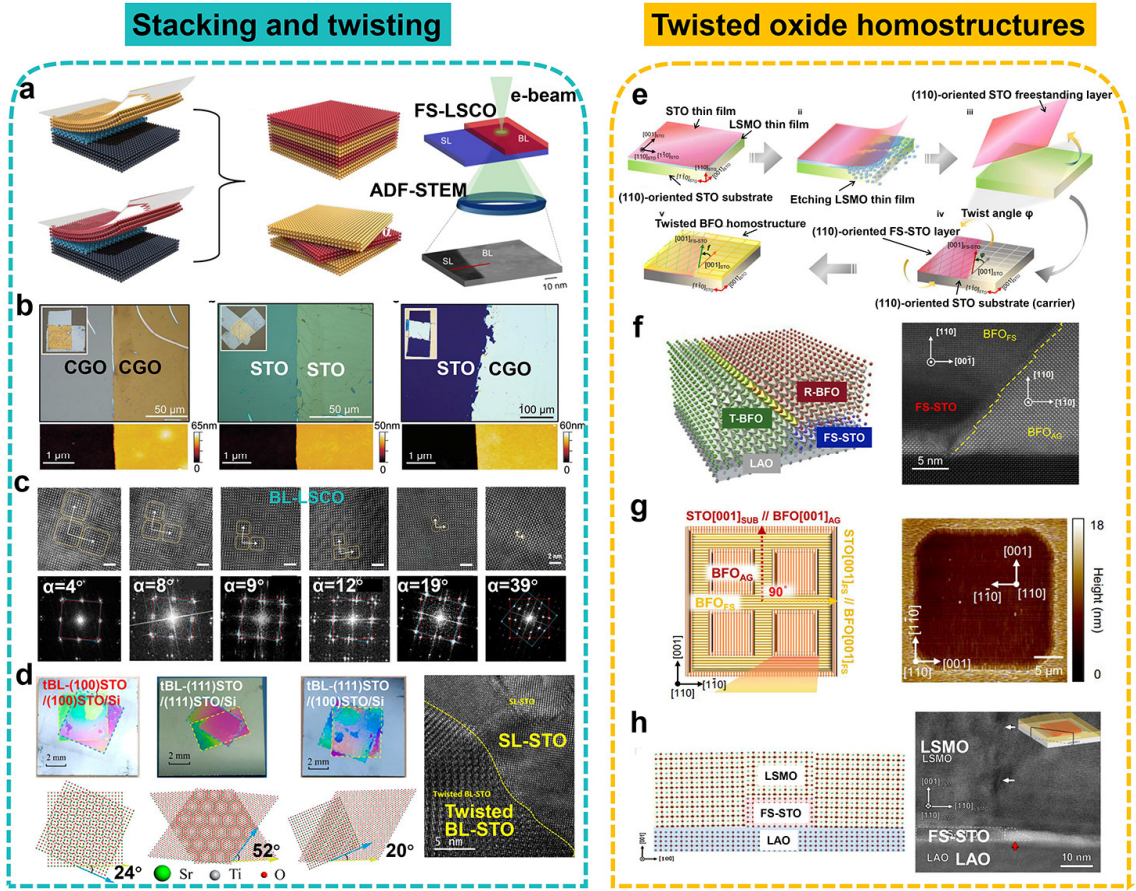

**Fig. 10. Controlled twistronics integration in freestanding thin films.**
**(a)** Schematic illustration of fabricating artificial heterointerface stacks via transfer of freestanding oxide thin films [26]. **(b)** Optical microscopy images of the bilayers of $Ce_{0.8}Gd_{0.2}O_{1.9}$ (40 nm)/$Ce_{0.8}Gd_{0.2}O_{1.9}$ (40 nm), $SrTiO_3$ (30 nm)/$SrTiO_3$ (30 nm) [26]. **(c)** Evolution of moiré patterns in twisted bilayer (BL) $La_{0.8}Sr_{0.2}CoO_3$. High-resolution STEM images and corresponding fast Fourier transform (FFT) images of BL freestanding $La_{0.8}Sr_{0.2}CoO_3$ thin films at varied twist angles ($\alpha$) [212]. **(d)** Optical images of twisted BL freestanding $SrTiO_3$ thin films on Si, simulated schematics with crystallographic orientations and high-resolution image of a boundary region revealing moiré patterns in the bilayer regime [213]. **(e)** Process flow for fabricating twisted lateral oxide homostructures [217]. **(f)** Designer lateral homostructure with coexisted tetragonal-like (T-$BiFeO_3$) and rhombohedral-like (R-BFO) $BiFeO_3$ phases. Corresponding HAADF image of the interface area observed along the 001 direction of $SrTiO_3$ substrate [217]. **(g)** Schematic of designer $BiFeO_3$ (110) lateral homostructures with orthogonally patterned stripes along $BiFeO_3$ 001.Topography of patterned lateral (110)-oriented $BiFeO_3$ homostructures [217]. **(h)** TEM image from the domain boundary region of $La_{0.7}Sr_{0.3}MnO_3$ homostructures [218].

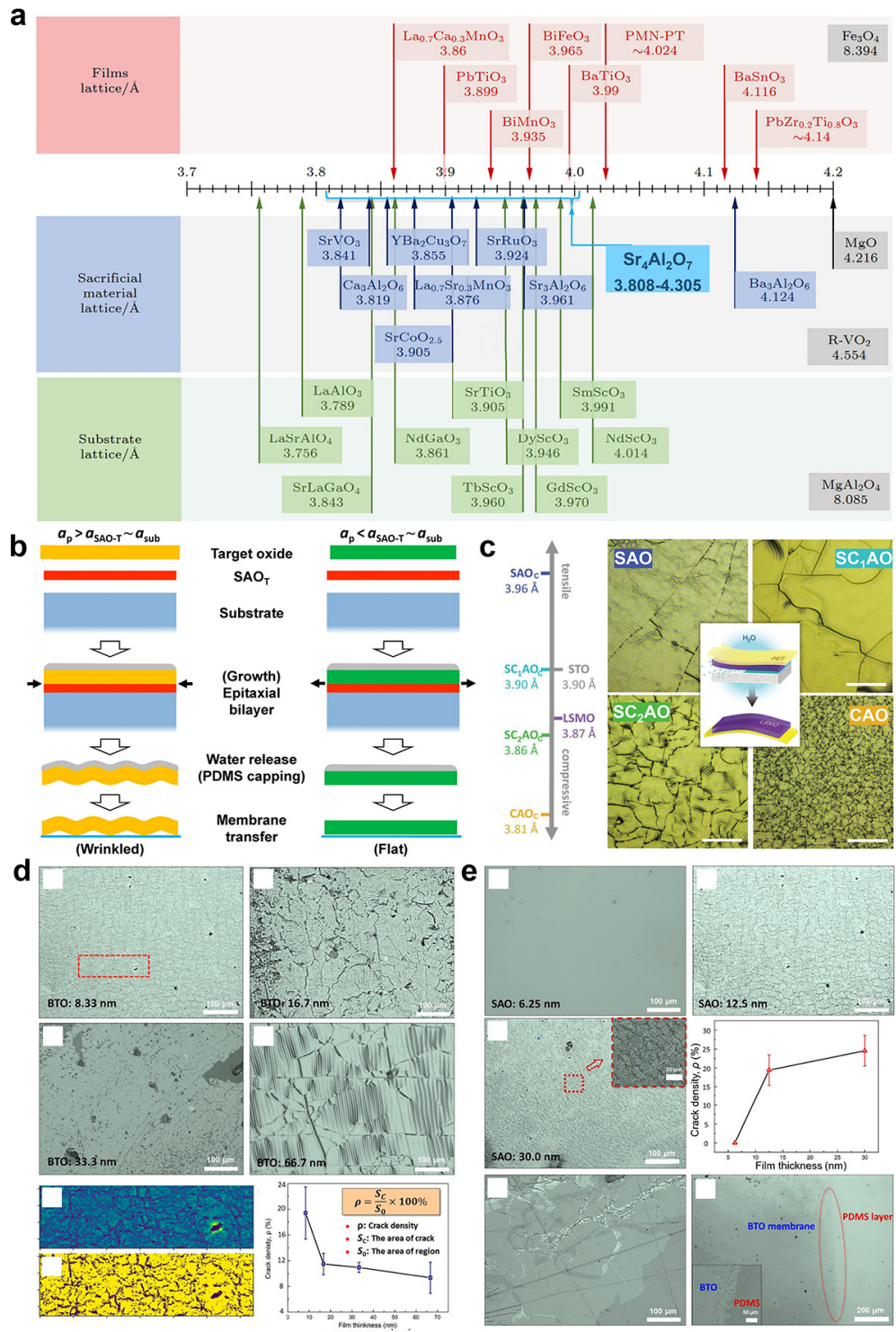

**Fig. 11. Structural integrity modulation.**
**(a)** Lattice parameters of materials employed in sacrificial etching technology [1]. **(b)** Schematic diagrams illustrating crack and wrinkle formation mechanisms in freestanding oxide thin films [75]. **(c)** Magnified images of the freestanding La$_{0.7}$Sr$_{0.3}$MnO$_3$ thin films on PET by etching Sr$_3$Al$_2$O$_6$, Sr$_2$CaAl$_2$O$_6$, SrCa$_2$Al$_2$O$_6$, and Ca$_3$Al$_2$O$_6$, respectively. Scale bar: 200 μm [175]. **(d)** Optical micrographs of freestanding BaTiO$_3$ thin films with varying thicknesses and corresponding crack density as a function of BaTiO$_3$ thickness. **(e)** Optical micrographs of 8.33-nm-thick freestanding BaTiO$_3$ thin films released from Sr$_3$Al$_2$O$_6$ sacrificial layers with different thicknesses and crack density of the BaTiO$_3$ thin films versus Sr$_3$Al$_2$O$_6$ thickness [189].

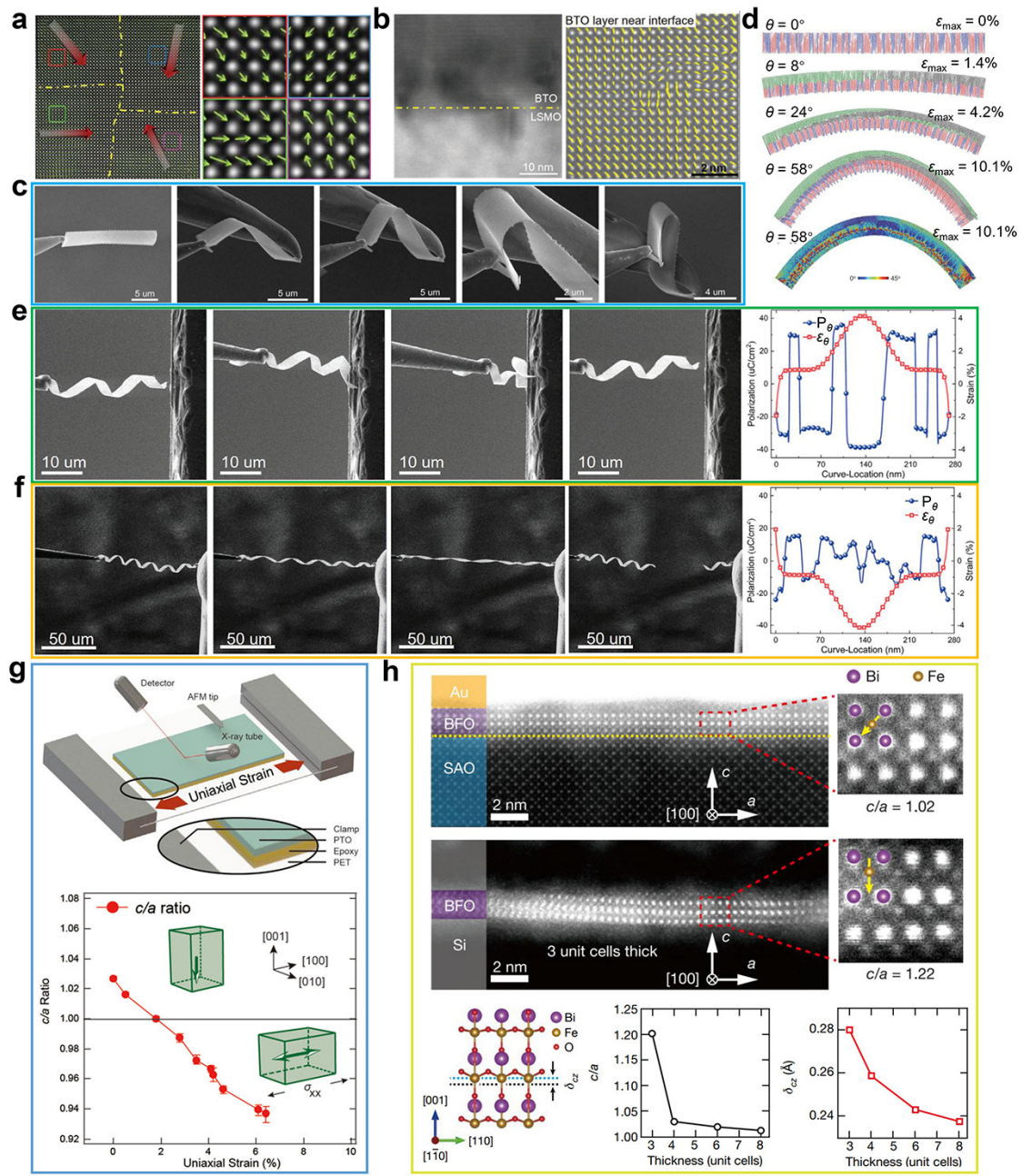

**Fig. 12. Strain-dominated mechanical responses of multifunctional freestanding oxide thin films.**
**(a)** Atomic-scale cross-sectional STEM together with Ti displacement of the multilayer BaTiO$_3$ region with strain$_{max}$ ~ 3.1%, and the magnified images of the four regions from the left. Dipole displacements are amplified by a factor of 15 for clarity [160]. **(b)** The high-resolution cross-sectional STEM-HAADF image of the La$_{0.7}$Sr$_{0.3}$MnO$_3$/BaTiO$_3$ interface (left) and the atomic displacement and polarization arrow map overlaid on STEM-HAADF image of the BaTiO$_3$ layer near the interface (right) [178]. **(c)** The series of in situ SEM images with the bending process of a BaTiO$_3$ nanobelt (20 mm by 4 mm by 120 nm) [160]. **(d)** Shape recovery of the BaTiO$_3$ membrane upon bending at 300 K. The bottom image is the spatial distribution of angle deviation $h$ with respect to the a or c nanodomain at $\theta = 58°$ [160]. **(e)** In situ SEM images of the La$_{0.7}$Sr$_{0.3}$MnO$_3$/BaTiO$_3$ nanosprings during the compressive process. And the corresponding polarization component $P_\theta$ and strain distribution $\varepsilon_\theta$ along the middle line of the outer surface under compress test from the phase-field simulations [178]. **(f)** In situ SEM images of the La$_{0.7}$Sr$_{0.3}$MnO$_3$/BaTiO$_3$ nanosprings during the elongation process [178]. **(g)** Schematic of stretching stage of freestanding PbTiO$_3$ thin film and the evolution of $c/a$ ratio with the uniaxial strain. Inset: representation of the as-grown and stretched polarization state (indicated by the arrow) [76]. **(h)** Cross-sectional HAADF images of a three-unit-cell BiFeO$_3$ film before (top) and after (middle) releasing the film, showing an R-like phase with polarization along the <111> direction and a T-like phase with polarization along the <001> direction, respectively. And the calculated giant polarization and lattice distortion in ultrathin freestanding BiFeO$_3$ films are shown in the bottom, in which the lower left corner is the structure of a three-unit-cell-thick BiFeO$_3$ film. The off-center displacement ($\delta_{cz}$) is defined as the distance along the out-of-plane direction between the centers of the neighboring Bi ions (dotted black line) and Fe ions (dotted blue line). The two at the bottom right are respectively the evolution of the average $c/a$ ratio (left) and $\delta_{cz}$ (right) as a function of thickness shows the increase of $c/a$ and the polarization in freestanding BiFeO$_3$ films when approaching the 2D limit [34].

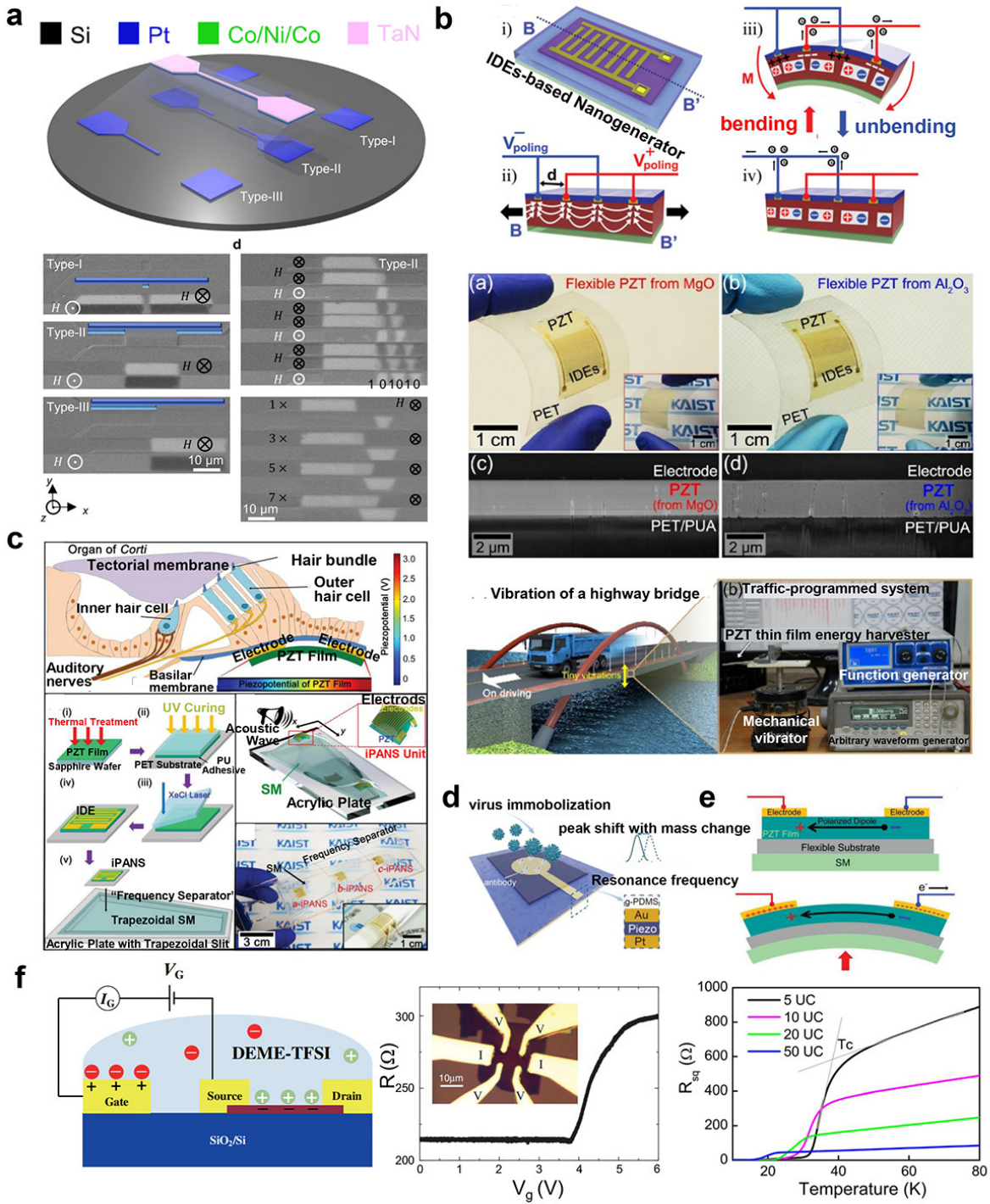

**Fig. 13. Device application of freestanding thin-film transfer and integration technology.**
**(a)** Local engineering of racetrack using pre-patterned Pt underlayers [190]. **(b)** Schematic of the working principle of a flexible freestanding PZT thin film nanogenerator (NG) based on interdigitated electrodes (IDEs) (top); freestanding PZT thin-film energy harvesters from MgO and Al₂O₃ wafers (middle); Programmable vibration system simulating traffic on a highway bridge with integrated energy harvester (bottom) [36,100]. **(c)** Integrated design of bioinspired artificial hair cells (AHCs) based on flexible freestanding PZT films: from mammalian cochlear model, laser lift-off fabrication, acoustoelectric conversion mechanism to experimental validation [35]. **(d)** Schematic of the piezoelectric freestanding thin-film based virus biosensor [55]. **(e)** Schematic of the inorganic-based piezoelectric acoustic nanosensor (iPANS) unit's piezoelectric generation principle: i) before and ii) after bending deformation [35]. **(f)** Schematic illustration of the FeSe thin flake electric-double-layer transistor (EDLT) device. Ionic liquid DEME-TFSI serves as the dielectric, covering the sample and gate electrodes (left). Gate voltage dependence of the resistance of an FeSe thin flake with a typical thickness of ~10 nm during a continuously swept positive gate voltage at a scan rate of 1 mV s$^{-1}$ (middle). Inset: Optical image of the EDLT device used in the present study. And transport measurement of ultrathin FeSe films on SrTiO$_3$ with varied thickness (right) [106,229].

# Freestanding Thin-Film Materials

## Graphical abstract

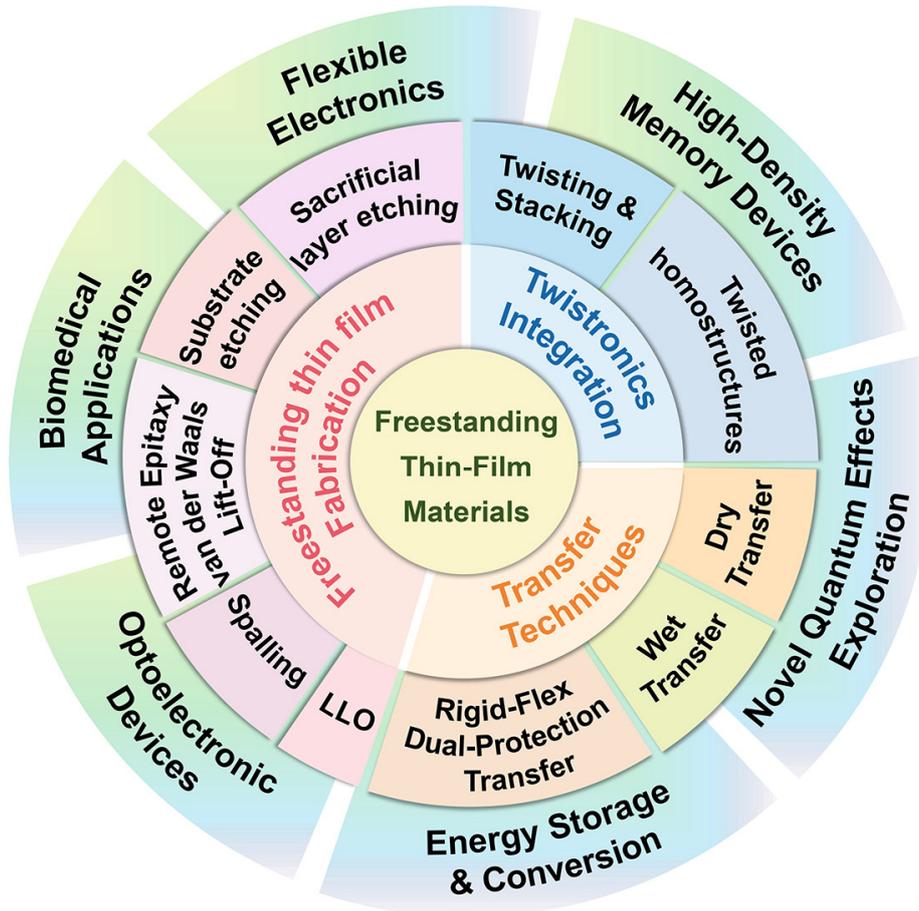

Freestanding thin-film materials overcome substrate clamping to unlock intrinsic material properties and enable cross-platform integration. This review details fabrication techniques (physical delamination, chemical etching) and transfer strategies for these films. It highlights their unique advantages in strain engineering, extreme mechanics, and interface decoupling, driving applications in flexible/sensing devices and quantum exploration. Challenges in scalability and stability are discussed, with promising future prospects.